\documentstyle[12pt]{article}
\newcommand{\GG}{$\propto\Gamma^{*}\Gamma\,$}
\newcommand{\G}{$\propto\Gamma(\Gamma^{*})\,$}
\newcommand{\Gp}{$\Gamma(\Gamma^{*}\,)$}
\newcommand{\GGp}{$\Gamma^{*}\Gamma\,$}

\newcommand{\bce}{\begin{center}}
\newcommand{\ece}{\end{center}}
\newcommand{\beq}{\begin{equation}}
\newcommand{\eeq}{\end{equation}}
\newcommand{\bea}{\vspace{0.25cm}\begin{eqnarray}}
\newcommand{\eea}{\end{eqnarray}}

\newcommand{\ba}{\begin{array}}
\newcommand{\ea}{\end{array}}

\newcommand{\doublespace}{
    \renewcommand{\baselinestretch}{1.6}\large\normalsize}

\def\lsim{\mathrel{\rlap{\lower4pt\hbox{\hskip1pt$\sim$}}
    \raise1pt\hbox{$<$}}}         
\def\gsim{\mathrel{\rlap{\lower4pt\hbox{\hskip1pt$\sim$}}
    \raise1pt\hbox{$>$}}}         

\def\beq{\begin{equation}}
\def\endeq{\end{equation}}
\def\bea{\begin{eqnarray}}
\def\arr{\begin{eqnarray}}
\def\endarr{\end{eqnarray}}

\makeindex
\begin{document}
\phantom{.}\hspace{9.0cm}{\large \bf KFA-IKP(Th)-1995-01}
\phantom{.}\hspace{9.5cm} January, 1995
\vspace{2cm}
\begin{center}
{\bf \huge Glauber theory of final-state interactions in
$(e,e'p)$ scattering\\}
\vspace{1cm}
{\bf N.N.Nikolaev$^{1,2)}$,
J.Speth$^{1)}$, B.G.Zakharov$^{2)}$ } \medskip\\
{\small \sl

$^{1)}$IKP(Theorie), Forschungszentrum  J\"ulich GmbH.,\\
D-52425 J\"ulich, Germany \\
$^{2)}$L.D.Landau Institute for Theoretical Physics, \\
GSP-1, 117940, ul.Kosygina 2, V-334 Moscow, Russia
\vspace{1cm}\\}
{\bf \LARGE A b s t r a c t \bigskip\\}
\end{center}
We develop the Glauber theory description of final-state
interaction (FSI) in quasielastic $A(e,e'p)$ scattering.
The important new effect is an interaction between the
two trajectories which enter the calculation of
FSI-distorted one-body density matrix and are connected
with incoherent elastic rescatterings of the struck proton
on spectator nucleons.
We demonstrate that FSI distortions of the missing
momentum distribution are large over the whole range
of missing momenta. Important finding is that incoherent
elastic rescatterings of the ejected proton lead
to a strong quantum mechanical distortions of both
the longitudinal and transverse missing momentum distributions.
It is shown that allowance for finite longitudinal size of the
interaction
region for proton-nucleon collision neglected in the
standard Glauber model drastically affects the theoretical
predictions at high longitudinal missing momentum.
We also find very large corrections to the missing momentum
distribution calculated within the local-density approximation.

\newpage
\doublespace

\section{Introduction}

In recent years much experimental and theoretical
efforts have been directed towards the investigation
of the final-state interaction (FSI) effects in
quasielastic $A(e,e^{\prime}p)$ scattering at high
$Q^{2}$. The quantity which is usually used to
characterize the strength of the FSI in a certain
kinematical region, $D$, of the missing energy , $E_{m}$,
and the missing momentum, $\vec{p}_{m}$,
is the nuclear transparency, $T_{A}(D)$, defined as
the ratio
\beq
T_{A}(D)=\frac{\int_{D} dE_{m}d^{3}\vec{p}_{m}
d\sigma (E_{m},\vec{p}_{m})}
{\int_{D} dE_{m}d^{3}\vec{p}_{m}
d\sigma_{PWIA}(E_{m},\vec{p}_{m})}\,.
\label{eq:1.1}
\eeq
Here $d\sigma (E_{m},\vec{p}_{m})$ is the experimentally
measured cross section, $d\sigma_{PWIA}(E_{m},\vec{p}_{m})$
is the theoretical cross section calculated in the plane
wave impulse approximation (PWIA) when FSI is not
taken into account. The missing energy is defined as
$E_{m}=m_{A}-m_{p}-E_{A-1}$, where $m_{A}$ and $m_{p}$
are the target nucleus and proton mass, respectively,
$E_{A-1}$ is the energy of the residual nucleus.
The missing momentum is connected with the virtual
photon three-momentum, $\vec{q}$, and the momentum
of the ejected proton, $\vec{p}$,
 $\vec{p}_{m}=\vec{q}-\vec{p}$.

The strong interactions that the stuck proton undergoes
 during its propagation
through the nuclear medium lead to the deviation
of $T_{A}$ from unity. It is expected \cite{M1,B1} however, that,
at asymptotically high $Q^{2}$, $T_{A}$ must tend
to unity due to the color transparency (CT)
phenomenon in QCD [3--5] (for the recent review on CT
see ref. \cite{NNZ}).
{}From the point of view of the multiple scattering
theory this effect corresponds to a cancellation
between the rescattering amplitudes with elastic (diagonal)
and inelastic (off-diagonal) intermediate states. These
coupled-channel effects only
become important at sufficiently high $Q^{2}$ and
the recent quantum mechanical analysis \cite{onsetCT}
of $A(e,e'p)$ scattering has
shown that CT effect from the off-diagonal
contribution to FSI
is still very small
in the region $Q^{2}\lsim 10$ GeV$^{2}$.
The energy ($Q^{2}$) dependence of the nuclear transparency
observed in the NE18 experiment \cite{NE18} is
in agreement with the one predicted in \cite{onsetCT}.
Thus, there are reasons to expect that in the region $Q^{2}\sim 2-10$
 GeV$^{2}$, which is particularly interesting from
the point of view of future high-statistics experiments at CEBAF,
FSI in $A(e,e'p)$ scattering will be dominated by
elastic rescatterings of
the struck proton on the spectator nucleons.
In this region of $Q^{2}$ the typical kinetic energy
of the struck proton $T_{kin}\approx Q^{2}/2m_{p}\gsim 1$
GeV is sufficiently large and FSI can be treated in the framework
of the standard
Glauber model \cite{Gl}.
The purpose of the present paper is the Glauber theory description
of the missing momentum distribution in inclusive
$(e,e'p)$. We focus on the region of missing momenta
$p_{m}\lsim k_{F}$ ( $k_{F}\sim 250$ MeV/c is the
Fermi momentum).
Such an analysis is interesting for several reasons:

First, understanding the $\vec{p}_{m}$-dependence
of FSI effects is necessary for
accurate interpretation of the
data from NE18 experiment \cite{NE18} and from future
experiments at CEBAF. The point is that experimentally one
measures the $A(e,e'p)$ cross section only in a certain
restricted window $D$ in the missing momentum.
Because  FSI affects the missing momentum
distribution as compared to the PWIA case, the
absolute value and the energy dependence of the experimentally
measured nuclear transparency will be different for different
kinematical domains $D$. Consequently, understanding the
$p_{m}$ and $Q^{2}$ dependence of the conventional FSI effects
is imperative for disentangling
the small CT effects at CEBAF and beyond.

Secondly, the
still another CT effect which can be obscured by FSI is an
asymmetry of nuclear transparency about $p_{m,z}=0$
\cite{JK,Jasym,Boffi1}
(as usual, we choose the $z$ axis
along the virtual photon's three-momentum). The CT induced
forward-backward asymmetry increases with $Q^{2}$.
However, similar
forward-backward asymmetry is generated by FSI
already at the level of elastic rescatterings
of the struck proton from the spectator nucleons. It is
a consequence of the nonzero real part of the elastic
$pN$ amplitude. The qualitative estimates \cite{NNZ} show that
at $Q^{2}\lsim 10$ GeV$^{2}$, the FSI-induced asymmetry can
overcome the CT-induced effect.
For this reason, the interpretation of results from
the future CEBAF experiments on the forward-backward
asymmetry as a signal for the onset of CT requires the accurate
calculation of the missing momentum distribution in the Glauber
model.

At last but not the least, the quantitative theory of FSI
in quasielastic $(e,e'p)$ scattering is interesting from
the point of view of the nuclear physics as well. At high
momenta, the single-particle momentum distribution (SPMD)
is sensitive to short
 range $NN$ correlations in nuclei. Because of FSI effects,
the experimentally measured missing momentum distribution in
inclusive
$A(e,e'p)$ scattering may considerably differ from the real SPMD.
The recent study \cite{He4,D2} has shown that even in light
nuclei ($D,^{4}He$), in which the probability of FSI is still
small, at high missing momenta
a nontrivial "interference" of the FSI effects and short range
$NN$ correlation takes place. In heavier nuclei, which we study
in the present paper, FSI effects turn out to be strong even
in the region $p_{m}\lsim 300$ MeV/c, where
the role of the short range $NN$ correlation is still marginal.
Nevertheless our results help to get insight into the role of the
FSI at the missing momenta $p_{m}\sim 300$ MeV/c which are close to
the kinematical region where the short range $NN$ correlations
become important. Our particularly important finding is that the
incoherent rescatterings of the struck proton from the spectator
nucleons considerably affect the longitudinal missing momentum
distribution as compared to SPMD. They
lead to large tails in the
missing momentum distribution at high $|p_{m,z}|$. The observed
effect is of purely quantum-mechanical origin and defies the
classical treatment.

One important finding from our study of FSI
a natural applicability limit for the Glauber formalism
in the case of $A(e,e'p)$ reaction. It is connected with the finite
longitudinal size of the interaction region for the proton-nucleon
collisions, which is about the proton radius and is neglected
in standard applications of the multiple scattering theory.
We show that for this reason
the standard Glauber model predicts an anomalously slow  decrease
( $\propto|p_{m,z}|^{-2}$) of the missing momentum
distribution at high longitudinal missing momenta.
The physical origin of this anomalous behavior is an
incorrect treatment in the Glauber model of the incoherent
rescatterings of the struck proton on the adjacent spectator
nucleons,
when the separation between the struck proton and spectator
nucleons is comparable with the proton size.
Our estimates show that there can be large uncertainties due
the finite-proton size effects at
$|p_{m,z}|\gsim 500$ MeV/c. In this region of $|p_{m,z}|$,
besides the short range $NN$ correlations,
the experimentally measured missing momentum distribution becomes
sensitive to the finite-proton size effects in FSI, which makes
the experimental study of $NN$ correlations much more difficult.
It is important that this novel sensitivity of the missing
momentum distribution at
high $|p_{m,z}|$ to the proton size does not disappear at high
$Q^{2}$, and that the same situation takes place for
the multiple scattering theory as a whole when inelastic
(off-diagonal)
rescatterings of the struck proton are included. For this
reason the effect must be taken into account in the interpretation
of the experimental data in high missing momentum region from future
experiments at large  $Q^{2}$.

Our paper is organized as follows. In section 2 we derive the
formulas for calculation of the missing momentum distribution
in quasielastic $(e,e'p)$ scattering within the Glauber model.
We discuss briefly the generalization of our formalism to
the case with inclusion of the CT effects as well. We conclude
section 2 with comments on other works on the application
of the Glauber model to $(e,e'p)$ reaction.
Section 3 is
devoted to the detailed comparison of the Glauber model with
the optical potential approach. In particular, we discuss
the difference between the treatment in these two models of the
contribution to the FSI from the incoherent rescatterings of
the struck proton in the nuclear medium. In conclusion of section 3
we discuss the formal analogy between the treatment of the FSI
effects in the optical potential approach and the
Glauber formalism in the case of exclusive $(e,e'p)$
scattering.
In section 4 we derive the multiple-scattering series for the
transverse missing momentum distribution. We show that $p_{\perp}$
distribution can be formally represented in a form when all the
quantum mechanical distortion effects are contained in the
local missing momentum distribution calculated without inclusion
of the incoherent rescatterings of the struck proton on the
spectator nucleons.
In section 5 we discuss in detail the influence of the FSI
effects upon the longitudinal missing momentum distribution.
We show that the incoherent rescatterings of the struck proton
on the spectator nucleons considerably affect the measured
in quasielastic $(e,e'p)$
scattering missing momentum distribution at high $|p_{m,z}|$.
The qualitative quantum mechanical analysis of this phenomenon
is presented. We conclude section 5 by discussion of the
applicability limits of the Glauber model.
In section 6 we present our numerical results.
The summary and conclusions are presented in section 7.

One remark on our terminology is in order: In the present paper
we consider $(e,e'p)$ reaction without production of new hadrons.
We will use the term "inclusive $(e,e'p)$ scattering" for
the processes in which the final state of the residual nucleus
is not observed. The term "exclusive $(e,e'p)$ scattering"
will be used for the processes in which there is only one
knocked out nucleon (the struck proton).

\section{FSI and the missing-momentum distribution
in Glauber formalism }
In the present paper we will restrict ourselves
to the case of the mass number of the target
nucleus $A\gg 1$. In this case, neglecting the
center of mass correlations, we can write the
reduced nuclear amplitude of the exclusive
process
 $e\,+\,A_{i}\,\rightarrow\,\,e'\,+\,(A-1)_{f}\,+\,p$
in the form
\beq
M_{f}=\int d^{3}\vec{r}_{1}...d^{3}\vec{r}_{A}
\Psi_{f}^{*}(\vec{r}_{2},...,\vec{r}_{A})
\Psi_{i}(\vec{r}_{1},...,\vec{r}_{A})
S(\vec{r}_{1},...\vec{r}_{A})\exp(i\vec{p}_{m}\vec{r}_{1})\,.
\label{eq:2.1}
\eeq
Here $\Psi_{i}$ and $\Psi_{f}$ are wave functions
of the target and residual nucleus, respectively. The nucleon "1" is
chosen to be the struck proton.
For the sake of brevity in Eq (\ref{eq:2.1}) and hereafter
the spin and isospin variables are suppressed.
The function $S(\vec{r}_{1},...,\vec{r}_{A})$ describes
the FSI of the struck proton in the nuclear medium.
In the Glauber approximation it is given by
\beq
S(\vec{r}_{1},...,\vec{r}_{A})=\prod\limits_{j=2}^{A}
\left[1-\theta(z_{j}-z_{1})
\Gamma(\vec{b}_{1}-\vec{b}_{j})\right]\,,
\label{eq:2.2}
\eeq
where $\vec{b}_{j}$ and $z_{j}$ are the transverse and
longitudinal coordinates of the nucleons
and  $\Gamma(\vec{b})$ is the familiar profile function
of the elastic proton-nucleon scattering.
We use for $\Gamma(\vec{b})$ the standard high-energy
parameterization
\beq
\Gamma(b)={\sigma_{tot}(pN) (1-i\alpha_{pN})\over 4\pi B_{pN}}
\exp\left[-{b^{2}\over 2B_{pN}}\right] \,.
\label{eq:2.3}
\eeq
Here $\alpha_{pN}$ is the ratio of the real to imaginary part
of the forward elastic $pN$ amplitude, $B_{pN}$ is the diffractive
slope describing the $t$ dependence of the elastic proton-nucleon
cross section
\beq
\frac{d\sigma_{el}(pN)}{dt}=
\left.\frac{d\sigma_{el}(pN)}{dt}\right|_{t=0}
\exp(-B_{pN}|t|)\,.
\label{eq:2.4}
\eeq
In the Glauber's high-energy approximation, the struck proton
propagates along the straight-path trajectory and
can interact with the spectator nucleon "j"
only provided that $z_{j}>z_{1}$, which is an origin of
the step-function $\theta(z_{j}-z_{1})$
in the FSI factor (\ref{eq:2.2}). Physically it implies that we
neglect the finite longitudinal size of the region where
the struck proton interacts with the spectator nucleon.
The consequences of this assumption
will be discussed in section 5.

Our aim is the calculation of the (inclusive) missing momentum
distribution, $w(\vec{p}_{m})$, or nuclear
transparency $T_{A}(\vec{p}_{m})$,
in the kinematical conditions when the events of the whole range of
$E_{m}$ are included (in a sense this case corresponds to the
experimental situation of "poor-energy resolution").
In this case the missing momentum distribution may be written as
\beq
w(\vec{p}_{m})=\frac{1}{(2\pi)^{3}}\int dE_{m}S(E_{m},\vec{p}_{m})\,,
\label{eq:2.5}
\eeq
where $S(E_{m},\vec{p}_{m})$ is the spectral function
taking into account the FSI of the struck proton
\beq
S(E_{m},\vec{p}_{m})=\sum_{f}|M_{f}(\vec{p}_{m})|^{2}
\delta(E_{m}+E_{A-1}(\vec{p}_{m})+m_{p}-m_{A})\,.
\label{eq:2.6}
\eeq
At high $Q^2$, the sum over all final states of the
residual nucleus required for the calculation of the missing
momentum distribution from Eqs. (\ref{eq:2.5}), (\ref{eq:2.6}) can be
performed making use of the closure relation
\beq
\sum_{f}\Psi_{f}(\vec{r}_{2}^{\,'},...,\vec{r}_{A}^{\,'})
\Psi^{*}_{f}(\vec{r}_{2},...,\vec{r}_{A})=
\prod\limits_{j=2}^{A}\delta(\vec{r}_{j}-\vec{r}_{j}^{\,'})
\label{eq:2.7}
\eeq
Employing  Eq. (\ref{eq:2.7}) makes it possible to
represent $w(\vec{p}_{m})$ in the following form
\bea
w(\vec{p}_{m})=\frac{1}{(2\pi)^{3}}\int
d^{3}\vec{r}_{1}d^{3}\vec{r}_{1}^{\,'}
\prod\limits_{j=2}^{A}d^{3}\vec{r}_{j}
\exp[i\vec{p}_{m}(\vec{r}_{1}-\vec{r}_{1}^{\,'})]
\,\,\,\,\,\,\nonumber\\ \times
\Psi_{i}(\vec{r}_{1},\vec{r}_{2},...,\vec{r}_{A})
\Psi_{i}^{*}(\vec{r}_{1}^{\,'},\vec{r}_{2},...,\vec{r}_{A})
S(\vec{r}_{1},\vec{r}_{2},...,\vec{r}_{A})
S^{*}(\vec{r}_{1}^{\,'},\vec{r}_{2},...,\vec{r}_{A})\,.
\label{eq:2.8}
\eea
Thus, from the point of view of the nuclear physics the calculation
of the missing momentum distribution in quasielastic $(e,e'p)$
scattering is reduced to the evaluation of the ground state
expectation value for a special many-body operator
\beq
w(\vec{p}_{m})=\langle\Psi_{i}|U(\vec{p}_{m})|
\Psi_{i}\rangle\,,
\label{eq:2.9}
\eeq
where
\beq
\langle \vec{r}_{1}^{\,'},...,\vec{r}_{A}^{\,'}|U(\vec{p}_{m})|
\vec{r}_{1},...,\vec{r}_{A}\rangle=\frac{1}{(2\pi)^{3}}
\exp[i\vec{p}_{m}(\vec{r}_{1}-\vec{r}_{1}^{\,'})]
\prod\limits_{j=2}^{A}\delta(\vec{r}_{j}^{\,'}-\vec{r}_{j})
S^{*}(\vec{r}_{1}^{\,'},...,\vec{r}_{A}^{\,'})
S(\vec{r}_{1},...,\vec{r}_{A})\,.
\label{eq:2.10}
\eeq
The peculiarity of the operator $U$ (\ref{eq:2.10}) is that it
distorts the target nucleus wave function in the variables
$\vec{r}_{2},...,\vec{r}_{A}$ only when some of $\vec{r}_{i}$
are close to at least one of the two straight-path
trajectories beginning from the points
$\vec{r}_{1}$ and $\vec{r}_{1}^{\,'}$, which
arise after taking the square of the reduced
nuclear matrix element (\ref{eq:2.1}). The crucial point of
the further analysis is that FSI generates short range
interaction
between these two trajectories, which will be one of the
main factors on the distortion of missing momentum distribution
as compared to SPMD.

It is worth noting at this point that up to now we did not use the
concrete form of the Glauber model attenuation factor
(\ref{eq:2.2}). It is
clear that generalization of Eq. (\ref{eq:2.8}) to the case when the
CT effects are included is reduced to the following replacement
\beq
S(\vec{r}_{1},\vec{r}_{2},...,\vec{r}_{A})\Rightarrow
\frac{\langle p|\hat{S}_{3q}(\vec{r}_{1},\vec{r}_{2},...,\vec{r}_{A})
|E\rangle}
{\langle p|E\rangle}\,.
\label{eq:2.11}
\eeq
Here $\hat{S}_{3q}(\vec{r}_{1},\vec{r}_{2},...,\vec{r}_{A})$ is an
operator describing the evolution of the three-quark wave
function of the struck proton during its propagation
in the nuclear medium, $|E\rangle$
is an three-quark wave function which describes the state of
the proton after absorption of the virtual photon at point
$\vec{r}_{1}$. In terms of the electromagnetic current operator
$\hat J_{em}$, the ejectile wave function is expressed as
\cite{Jasym}
\beq
|E\rangle=\hat{J}_{em}(Q)|p\rangle=\sum\limits_{i}
|i\rangle\langle i|J_{em}(Q)|p\rangle=\sum\limits_{i}
G_{ip}(Q)|i\rangle\,,
\label{eq:2.12}
\eeq
where $G_{ip}(Q)=\langle i|J_{em}(Q)|p\rangle$ includes the
electromagnetic form factor of the proton as well as all
transition form factors for the electroexcitation of the proton
$e~+~p~\rightarrow~e'~+~i$. In the case of the nonrelativistic
oscillator quark model the evolution operator $S_{3q}$ can be
computed using the path integral technique \cite{S3q1,S3q2,S3q3}.
It is possible to evaluate this operator in the multiple
scattering approach as well \cite{onsetCT}. In the present paper
we restrict ourselves to the calculation of the missing momentum
distribution in the Glauber approximation. The analysis
taking into account the CT effects will be presented
elsewhere.

In the above analysis we factored out the electromagnetic current
matrix elements of the struck proton. In doing so, we
followed the usual tradition \cite{Boffi2} of neglecting
possible departures of these matrix elements from the PWIA
matrix elements, which may emerge because of the off-mass shell
effects \cite{Forest} induced by FSI.
Under this assumption the (local)
nuclear transparency as a function
of the missing momentum can be written as
\beq
T_{A}(\vec{p}_{m})=w(\vec{p}_{m})/n_{F}(\vec{p}_{m})\,,
\label{eq:2.13}
\eeq
where $n_{F}(\vec{p})$ is the SPMD
that can be expressed through the one-body nuclear
density matrix, $\rho(\vec{r},\vec{r}^{\,'})$,
\beq
n_{F}(\vec{p})=\frac{1}{(2\pi)^{3}}\int
d^{3}\vec{r}d^{3}\vec{r}^{\,'}\rho(\vec{r},\vec{r}^{\,'})
\exp[i\vec{p}(\vec{r}-\vec{r}^{\,'})]\,.
\label{eq:2.14}
\eeq
The normalization of $n_{F}(\vec{p})$ is as follows
$$\int d^{3}\vec{p}n_{F}(\vec{p})=1\,.$$

Integration of the missing momentum distribution $w(\vec{p}_{m})$
over the whole region of $\vec{p}_{m}$
gives the integral nuclear transparency
\beq
\int d^{3}\vec{p}_{m}w(\vec{p}_{m})=T_{A}\,.
\label{eq:2.15}
\eeq
Using Eqs. (\ref{eq:2.8}), (\ref{eq:2.15}) one can obtain
the following expression for $T_{A}$
\beq
T_{A}=\int \prod\limits_{j=1}^{A}d^{3}\vec{r}_{j}
|\Psi_{i}(\vec{r}_{1},...,\vec{r}_{A})|^{2}
|S(\vec{r}_{1},...,\vec{r}_{A})|^{2}\,.
\label{eq:2.16}
\eeq
Besides the distribution $w(\vec{p}_{m})$, which is normalized
to $T_{A}$, we will use in the present paper the missing momentum
distribution $n_{eff}(\vec{p}_{m})$,
which is normalized to unity
\beq
n_{eff}(\vec{p}_{m})=w(\vec{p}_{m})/T_{A}\,.
\label{eq:2.17}
\eeq

Equations (\ref{eq:2.2}), (\ref{eq:2.8}) can be used
as a basis for the calculation of the nuclear transparency
in reaction $(e,e'p)$ within the Glauber model.
Evidently, even without taking into account the
CT effects, evaluation of the nuclear transparency
(especially if we are interested in the $\vec{p}_{m}$
dependence of $T_{A}(\vec{p}_{m})$ or $w(\vec{p}_{m})$) is quite
an involved problem.
In this communication we confine ourselves
to an evaluation of FSI effects at moderate missing
momenta $p_{m} \lsim k_{F}$, where the simple
shell model is well known to give good description of SPMD and
short-range $NN$ correlations effects are marginal \cite{NNcor}.
In the opposite to that, at high missing
momenta $p_{m} \gsim k_{F}$, there emerges a
complicated pattern of the interference effects
between the short range $NN$ correlations in the nucleus
wave function \cite{NN1,NN2,NNcor}
and FSI of the struck proton
\cite{He4,D2}. Furthermore, even at large $p_{m}$, FSI effects
turn out to be numerically substantially
larger than $NN$ correlation effects
\cite{He4,D2,Jresc}.
Therefore, as far as the salient features of
FSI, in particular the understanding of the r\^ole of
interaction between the two trajectories in FSI factor,
are concerned, it is reasonable to
use a simple independent particle nuclear shell model for calculation
of the missing momentum distribution in the region
$p_{m}\lsim k_{F}$.

In this case,
making use of the Slater determinant form of the shell model
wave function we can write the product of the wave functions
appearing in Eq. (\ref{eq:2.8}) in the form
\bea
\Psi_{i}(\vec{r}_{1},\vec{r}_{2},...,\vec{r}_{A})
\Psi_{i}^{*}(\vec{r}_{1}^{\,'},\vec{r}_{2},...,\vec{r}_{A})=
\frac{1}{Z}\sum\limits_{n}
\phi_{n}(\vec{r_{1}})\phi_{n}^{*}(\vec{r}_{1}^{\,'})
\rho_{n}(\vec{r}_{2},...,\vec{r}_{A})\nonumber\\
+\frac{1}{Z}\sum\limits_{n\not=m}\phi_{n}(\vec{r}_{1})
\phi_{m}(\vec{r}_{1}^{\,'})
\rho_{n m}(\vec{r}_{2},...,\vec{r}_{A})\,.
\label{eq:2.181}
\eea
Here  $Z$ is the number of protons in the target nucleus,
$\phi_{n}(\vec{r})$ are single-particle shell model proton
wave functions. In the right hand side of Eq. (\ref{eq:2.181}) we
separated the sum over the struck proton states into the
diagonal part (the first term) and nondiagonal one (the second
term). The diagonal, $\rho_{n}(\vec{r}_{2},...,\vec{r}_{A})$, and
nondiagonal, $\rho_{n m}(\vec{r}_{2},...,\vec{r}_{A})$, $(A-1)$-body
distributions introduced in Eq. (\ref{eq:2.181}) are given by
\beq
\rho_{n}(\vec{r}_{2},...,\vec{r}_{A})=
\Psi_{i,n}(\vec{r}_{2},...,\vec{r}_{A})
\Psi_{i,n}^{*}(\vec{r}_{2},...,\vec{r}_{A})\,,
\label{eq:2.182}
\eeq
\beq
\rho_{n m}(\vec{r}_{2},...,\vec{r}_{A})=
\Psi_{i,n}(\vec{r}_{2},...,\vec{r}_{A})
\Psi_{i,m}^{*}(\vec{r}_{2},...,\vec{r}_{A})\,,
\label{eq:2.183}
\eeq
where $\Psi_{i,n}$ is
the $(A-1)$-body wave function which describes the system of $(A-1)$
nucleons obtained after removing of the proton in the state $n$ from
the target nucleus. The $(A-1)$-body nuclear density of this system,
given by the function $\rho_{n}$, is normalized to unity
\beq
\int \prod\limits_{j=2}^{A}d^{3}\vec{r}_{j}
\rho_{n}(\vec{r}_{2},...,\vec{r}_{A})=1\,.
\label{eq:2.184}
\eeq
In the case of the nondiagonal distribution $\rho_{n m}$
(\ref{eq:2.183})  the
corresponding integral over the coordinates $2-A$ is equal zero.
However, due to the presence of the Glauber
attenuation factors,
the contribution to the
integral over the coordinates of
the spectator nucleons in Eq. (\ref{eq:2.8}), related to the last term
in the product of the wave function (\ref{eq:2.181}), does not vanish.
None the less, making use of
the random phase approximation one can show that the
contribution to the missing momentum distribution
related to the nondiagonal part of the product of
the target nucleus wave functions is suppressed
by the factor $1/A$
in a comparison with the one from the first term in
Eq. (\ref{eq:2.181}). Thus, to the
leading order in $1/A$, the calculation of the missing momentum
distribution can be performed keeping in Eq. (\ref{eq:2.181})
only the diagonal term.

To proceed with the calculation of the missing momentum
distribution we will neglect influence of the Fermi correlations
on the $(A-1)$-body nuclear density $\rho_{n}$,
and
approximate it by
the factored form (notice that to the leading order in $1/A$
the dependence of $\rho_{n}$ on index $n$ can be neglected as well)
\beq
\rho_{n}(\vec{r}_{2},...,\vec{r}_{A})\approx
\prod\limits_{j=2}^{A}\rho_{A}(\vec{r}_{j})\,,
\label{eq:2.185}
\eeq
where $\rho_{A}(\vec{r})$ is the nucleon nuclear density normalized as
$$\int d^{3}\vec{r}\rho_{A}(\vec{r})=1\,.$$
The fact that the factored approximation for the many-body
nuclear density is a very good one for the purpose of the
calculation of the Glauber model attenuation factor for the
case of soft hadron nucleus scattering is
well known for a long time (for an extensive review on $hA$
scattering see \cite{Alkhaz}, the
correlation effect in the integrated nuclear transparency $T_{A}$
was discussed in \cite{Benhar,Jcor}).

Making use of Eqs. (\ref{eq:2.181}),  (\ref{eq:2.185})
we can represent the missing
momentum distribution (\ref{eq:2.8}) in the form
\beq
w(\vec{p}_{m})=\frac{1}{(2\pi)^{3}}\int
d^{3}\vec{r}_{1}d^{3}\vec{r}_{1}^{\,'}
\rho(\vec{r}_{1},\vec{r}_{1}^{\,'})
\Phi(\vec{r}_{1},\vec{r}_{1}^{\,'})
\exp[i\vec{p}_{m}(\vec{r}_{1}-\vec{r}_{1}^{\,'})]\,,
\label{eq:2.20}
\eeq
which only differs from the SPMD for the presence of the FSI factor
\beq
\Phi(\vec{r}_{1},\vec{r}_{1}^{\,'})=
\int \prod\limits_{j=2}^{A}\rho_{A}(\vec{r}_{j}) d^{3}\vec{r}_{j}
S(\vec{r}_{1},\vec{r}_{2},...,\vec{r}_{A})
S^{*}(\vec{r}_{1}^{\,'},\vec{r}_{2},...,\vec{r}_{A})\,,
\label{eq:2.21}
\eeq
which describes the FSI-distortion of
the one-body shell model
proton density matrix
\beq
\rho(\vec{r},\vec{r}^{\,'})=\frac{1}{Z}\sum\limits_{n}
\phi_{n}(\vec{r})\phi_{n}^{*}(\vec{r}^{\,'})\,.
\label{eq:2.19}
\eeq

Eq. (\ref{eq:2.20}) is a counterpart of the conventional
formula (\ref{eq:2.14}) for SPMD.
Due to the dependence of FSI factor
$\Phi(\vec{r}_{1},\vec{r}_{1}^{\,'})$ on $\vec{r}_{1}$ and
$\vec{r}^{\,'}_{1}$, the normalized missing momentum distribution
(\ref{eq:2.17}) does not coincide with the Fermi distribution.
The
FSI factor (\ref{eq:2.21}) can not be represented in a
factored
form in the variables $\vec{r}_{1}$ and $\vec{r}_{1}^{\,'}$.
In particular, because of FSI
the function $w(\vec{p}_{m})$ is not isotropic one.
We remind that our approach only
is applicable in the region of relatively small missing
momenta $p_{m}\lsim k_{F}$. We postpone the analysis of the high
missing momentum region for further publications.

By using formula (\ref{eq:2.2}) we obtain the closed analytical
expression for FSI factor
(\ref{eq:2.21})
\bea
\Phi(\vec{r}_{1},\vec{r}_{1}^{\,'})=\left[1-\frac{1}{A}
\int
d^{2}\vec{b}\Gamma(\vec{b}_{1}-\vec{b})t(\vec{b},z_{1})-\frac{1}{A}
\int
d^{2}\vec{b}\Gamma^{*}(\vec{b}_{1}^{\,'}-\vec{b})t(\vec{b},z_{1}^{'})
\right.\nonumber\\ \left. +\frac{1}{A}
\int d^{2}\vec{b}
\Gamma^{*}(\vec{b}_{1}^{\,'}-\vec{b})\Gamma(\vec{b}_{1}-\vec{b})
t(\vec{b},max(z_{1},z_{1}^{\,'})\right]^{A-1}\,,
\label{eq:2.22}
\eea
here we introduced the partial optical thickness function
\beq
t(\vec{b},z)=A\,\int\limits_{z}^{\infty}d\xi
\rho_{A}(\vec{b},\xi)\,.
\label{eq:2.23}
\eeq
The FSI factor can further be simplified
exploiting the fact that the
partial optical thickness $t(\vec{b},z)$ is a smooth
function of the impact parameter $\vec{b}$
as compared to the nuclear profile function $\Gamma(\vec{b})$. Then,
to the zeroth order in the small parameter $B_{pN}/R_{A}^{2}$ (
$R_{A}$
is the nucleus radius)  the result of integration over the impact
parameter $\vec{b}$ in the terms \G in Eq. (\ref{eq:2.22}) is given by
\beq
\int d^{2}\vec{b}\Gamma(\vec{b}_{1}-\vec{b})t(\vec{b},z_{1})\approx
\frac{\sigma_{tot}(pN)(1-i\alpha_{pN})}{2}t(\vec{b},z)\,.
\label{eq:2.24}
\eeq
In the same
approximation for the term \GG in the brackets in Eq. (\ref{eq:2.22})
we have
\beq
\int d^{2}\vec{b}
\Gamma^{*}(\vec{b}_{1}^{\,'}-\vec{b})\Gamma(\vec{b}_{1}-\vec{b})
t(\vec{b},z)\approx\eta(\vec{b}_{1}-\vec{b}_{1}^{\,'})\sigma_{el}(pN)
t(\frac{1}{2}(\vec{b}_{1}+\vec{b}_{1}^{\,'}),z)\,,
\label{eq:2.25}
\eeq
where the function $\eta(\vec{b})$ is given by
\bea
\eta(\vec{b})=\frac{\int
d^{2}\vec{\Delta}\Gamma^{*}(\vec{b}-\vec{\Delta})
\Gamma(\vec{\Delta})}
{\int d^{2}\vec{\Delta} |\Gamma(\vec{\Delta})|^{2}}\;\;\;\;\nonumber\\
=\frac{1}{\pi\sigma_{el}(pN)}\int d^{2}\vec{q}\frac{d\sigma_{el}(pN)}
{dq^{2}}\exp(i\vec{q}\vec{b})
=\exp\left[-\frac{\vec{b}^{\,2}}{4B_{pN}}\right]
\label{eq:2.26}
\eea
(
We checked that in the case of the nuclear mass number
$A\gsim 10$ the corrections to the Eqs. (\ref{eq:2.24}),
(\ref{eq:2.25}) connected with the neglected higher order terms in
the ratio $B_{pN}/R_{A}^{2}$ lead to the corrections in the
final numerical predictions for the missing momentum
distribution and $T_{A}$ which do not exceed 1-3\%.)
Finally,
making use of Eqs. (\ref{eq:2.22}), (\ref{eq:2.24}), (\ref{eq:2.25})
and exponentiating which is a good approximation at $A\gg 1$,
we arrive at the following
expression for the FSI factor (\ref{eq:2.21})
\bea
\Phi(\vec{r}_{1},\vec{r}_{1}^{\,'})=\exp\left[
-\frac{1}{2}\sigma_{tot}(pN)(1-i\alpha_{pN})t(\vec{b}_{1},z_{1})
-\frac{1}{2}\sigma_{tot}(pN)(1+i\alpha_{pN})t(\vec{b}_{1}^{\,'},z_{1}^
{'}) \right.\nonumber\\  \left.
+\eta(\vec{b}_{1}-\vec{b}_{1}^{\,'})\sigma_{el}(pN)
t(\frac{1}{2}(\vec{b}_{1}+\vec{b}_{1}^{\,'}),max(z_{1},z_{1}^{'}))
\right]\,.
\label{eq:2.27}
\eea
Below we will refer to the first two terms in the exponent in Eq.
(\ref{eq:2.27}) as \Gp terms , and the last one as \GGp term.
Notice that, were it not for the \GGp terms in the exponent, the
FSI factor would have factored into the two independent
attenuation factors which only depend on $\vec{r}$ and
$\vec{r}\,'$, respectively. The \GGp term
introduces an interaction between the two trajectories, which
is a steep function of $|\vec{b}_{1}-\vec{b}_{1}'|$. This
interaction substantially affects the observed missing momentum
distribution and shall be of major concern in this paper.

Besides the three-dimensional distribution $w(\vec{p}_{m})$
we will consider $p_{m,z}$ integrated $\vec{p}_{m\perp}$ distribution,
$w_{\perp}(\vec{p}_{m\perp})$, and $\vec{p}_{m\perp}$
integrated $p_{m,z}$ distribution, $w_{z}(p_{m,z})$.
Performing the integration of the distribution $w(\vec{p}_{m})$
given by Eq. (\ref{eq:2.20}) over transverse and longitudinal
component of the missing momentum one can obtain for
$p_{m,z}$ and $p_{m\perp}$ distributions
\beq
w_{z}(p_{m,z})=\frac{1}{2\pi}
\int d^{2}\vec{b} dz dz^{'}\rho(\vec{b},z,\vec{b},z^{'})
\Phi_{z}(\vec{b},z,z^{'})
\exp[i\vec{p}_{m,z}(z-z^{'})]\,,
\label{eq:2.28}
\eeq

\beq
w_{\perp}(\vec{p}_{m\perp})=\frac{1}{(2\pi)^{2}}
\int d^{2}\vec{b} d^{2}\vec{b}^{\,'}dz\rho(\vec{b},z,\vec{b}^{\,'},z)
\Phi_{\perp}(\vec{b},\vec{b}^{\,'},z)
\exp[i\vec{p}_{m\perp}(\vec{b}-\vec{b}^{\,'})]\,,
\label{eq:2.29}
\eeq
where
\bea
\Phi_{z}(\vec{b},z,z^{'})=\exp\left[
-\frac{1}{2}\sigma_{tot}(pN)(1-i\alpha_{pN})t(\vec{b},z)
-\frac{1}{2}\sigma_{tot}(pN)(1+i\alpha_{pN})t(\vec{b},z^{'})
\right.\nonumber\\
+\left.\sigma_{el}(pN)t(\vec{b},max(z,z^{'})\right]\,,
\label{eq:2.30}
\eea
\bea
\Phi_{\perp}(\vec{b},\vec{b}^{\,'},z)=\exp\left[
-\frac{1}{2}\sigma_{tot}(pN)(1-i\alpha_{pN})t(\vec{b},z)
-\frac{1}{2}\sigma_{tot}(pN)(1+i\alpha_{pN})t(\vec{b}^{\,'},z)
\right.\nonumber\\
+\left.\eta(\vec{b}-\vec{b}^{\,'})\sigma_{el}(pN)
t(\frac{1}{2}(\vec{b}+\vec{b}^{\,'}),z)\right]\,.
\label{eq:2.31}
\eea
Eqs. (\ref{eq:2.20}), (\ref{eq:2.27}),
(\ref{eq:2.28})--(\ref{eq:2.31}) form a basis for our evaluations
of the missing momentum distribution ( and $T_{A}(\vec{p}_{m})$) in
quasielastic $(e,e'p)$ scattering in the Glauber model.
The three-dimensional distribution (\ref{eq:2.20}), as it was mentioned
in section 1, is particularly
interesting from the point of view of using it
for an accurate comparison of the theoretical
predictions with experimental data on the nuclear transparency
obtained for a certain kinematical domain $D$, when the
nuclear transparency $T_{A}(D)$ is defined according to Eq.
(\ref{eq:1.1}).
In the present paper we will make for the first time such a comparison
of the Glauber model predictions with the data from the NE18
experiment \cite{NE18}.

The formalism presented in this section is rather simple and is
based
upon the same ideas as the original Glauber approach to the
hadron-nucleus interactions at small momentum transfer \cite{Gl}.
We gave a very detailed derivation mostly for the reason
that in the current literature there exist
discussions of the missing momentum distribution
in reaction $(e,e'p)$ within the same
Glauber model, which incorrectly treat
the important effect of interaction
between the two trajectories \cite{KYS1,RJ}.
For instance, the authors of ref.\cite{KYS1} neglect the
dependence on the
variable $\Delta\vec{r}_{1}=(\vec{r}_{1}-\vec{r}_{1}^{\,'})$ in
their counterpart of our FSI factor (\ref{eq:2.27}) and put in it
\beq
\vec{r}_{1}=\vec{r}_{1}^{\,'}=(\vec{r}_{1}+\vec{r}_{1}^{\,'})/2\,.
\label{eq:2.33}
\eeq
Doing so, they completely
missed the rapid dependence on $\Delta \vec{r}_{1}$
of the \GGp term in the exponent of FSI factor. Further,
using the Negele-Vautherin local density approximation (LDA)
\cite{LDA} for one-body density matrix (we comment more on this
approximation below)
\beq
\rho(\vec{r},\vec{r}^{\,'})=\rho_{A}
(\frac{1}{2}(\vec{r}+\vec{r}^{\,'}))W(\vec{r}-\vec{r}^{\,'})
\label{eq:2.32}
\eeq
(here $W(\vec{r}-\vec{r}^{\,'})$ is the Fourier transform of
the Fermi distribution), Kohama et al. find the
missing momentum distribution which is proportional to SPMD
Fermi momentum distribution. It is clear that in doing so they
missed all the distortion effects which, as we shall demonstrate
below, are quite strong. The same criticism is relevant to an
 analysis \cite{KYS2}
of quasielastic $(p,2p)$ scattering.

The fact that neglecting \cite{KYS1} the dependence of the absorption
factor
on the variable $\Delta\vec{r}_{1}$
is illegitimate
was noticed in \cite{RJ}.  None the less the authors
of ref. \cite{RJ} did not accomplish a complete analysis
of distortion effects in
$A(e,e'p)$ scattering taking into account
the dependence of the FSI factor
on $\Delta\vec{r}_{1}$. They restricted themselves
to accounting for the dependence of the FSI on the longitudinal
component of the vector $\Delta\vec{r}_{1}$ and put in the FSI
factor
\beq
\vec{b}_{1}=\vec{b}_{1}^{\,'}=(\vec{b}_{1}+\vec{b}_{1}^{\,'})/2\,.
\label{eq:2.34}
\eeq
For the reason that interaction between the two trajectories
is a steep function of
$\Delta\vec{b}_{1}=(\vec{b}_{1}-\vec{b}_{1}^{\,'})$, the
approximation (\ref{eq:2.34}) can not be justified.
Evidently, (\ref{eq:2.34}) precludes an accurate treatment
of the transverse missing momentum distribution. Ref. \cite{RJ}
also used LDA for the one-body density matrix.
It is easy to show using Eqs. (\ref{eq:2.20}),
(\ref{eq:2.27}) that the approximation (\ref{eq:2.34})
leads to the $p_{m,z}$
integrated
transverse missing momentum distribution which is proportional
to the $p_{z}$ integrated transverse SPMD. Furthermore,
in the region $p_{m}\lsim k_{F}$, where $n_{f}(p_{m})$ may be
approximated by the Gaussian form, in the resulting three-dimensional
missing momentum distribution the dependencies on the
transverse and longitudinal components will factorize with
having the same $p_{m\perp}$ dependence as the SPMD.
Such a factorization can not be correct, because
the term \GG in Eq. (\ref{eq:2.27}), which has the most
steep dependence on $\Delta\vec{b}_{1}$, and the terms
\G, which are smooth function of $\Delta\vec{b}_{1}$, have quite
  different dependence on $(z_{1}-z_{1}^{'})$.
Our numerical results  show that
three-dimensional missing momentum distribution (\ref{eq:2.20})
actually has a manifestly non-factorizable form. Therefore the
application of
the approximation (\ref{eq:2.34}) for the evaluation
of the three-dimensional missing momentum distribution
in $(e,e'p)$ scattering can not be justified.

Concluding discussion of the approaches of ref. \cite{KYS1,RJ}
one remark is in order on the LDA (\ref{eq:2.32}) for one-body
density matrix that was used in \cite{KYS1,RJ}. The LDA is widely
believed to be a very good approximation for heavy nuclei.
Our numerical results show
that in the case of calculation the missing momentum distribution
in $(e,e'p)$ scattering LDA turns out to be quite a crude
approximation
even for the nucleus mass number $A=40$. The comparison of
the results obtained using the full shell model density matrix
(\ref{eq:2.19})
and its parameterization in a factorizable form (\ref{eq:2.32}) will
be presented in section  6.

\section{Connection between
Glauber model and optical potential approach}

A comparison between the Glauber formalism
set up in section 2 and the optical potential approach (the
conventional distorted-wave impulse approximation  (DWIA))
that is usually used to describe  the
FSI effects in $(e,e'p)$ scattering at low $Q^{2}$ ( for the recent
review see ref. \cite{Boffi2}), is in order.
In the DWIA the FSI effects are taken into account by introducing
a phenomenological optical potential, $V_{opt}(\vec{r})$. Then the
distortion of the outgoing proton plane wave arises as a consequence
of the eikonal phase factor
\beq
S_{opt}(\vec{r})=
\exp\left[-\frac{i}{v}\int\limits_{z}^{\infty}d\xi
V_{opt}(\vec{b},\xi)\right]
\label{eq:3.1}
\eeq
( $v$ is the velocity of the struck proton).
The missing momentum distribution in this approach is given by Eq.
(\ref{eq:2.20}) with the following  factored FSI factor
\beq
\Phi_{opt}(\vec{r}_{1},\vec{r}_{1}^{\,'})=
S_{opt}(\vec{r}_{1}) S_{opt}^{*}(\vec{r}_{1}^{\,'})\,,
\label{eq:3.2}
\eeq

The important feature of the DWIA
is that optical potential does not depend on the individual
coordinates of the spectator nucleons. Thus the optical potential
in the DWIA embodies an effective description of the influence
of the nuclear medium upon the wave function of the struck proton.

As a matter of fact, the Glauber model attenuation factor
(\ref{eq:2.2}) is but the embodiment of solving
of the wave equation for the wave function of the struck proton
in the eikonal approximation as well. Nevertheless there is
an important conceptual difference between the DWIA and
the Glauber model approach. In the DWIA the FSI effects are
taken into account at the level of the wave function of the
ejected proton.
However it is clear that a rigorous evaluation of the
probability distribution for a subsystem (the struck proton
in the case under consideration) for the process including a complex
system (the system of the struck proton and spectator nucleons
in our case) requires calculations at the level of the
subsystem density matrix.
Eqs. (\ref{eq:2.2}), (\ref{eq:2.8}) embody precisely this
procedure in the case of quasielastic $(e,e'p)$
scattering. Indeed, as was mentioned above the Glauber model
attenuation factor (\ref{eq:2.2}) realizes the solving of the wave
equation for
the outgoing proton wave function for a certain configuration of the
spectator nucleons, $\tau=\{\vec{r}_{2},...,\vec{r}_{A}\}$.
As usual we assume that at high energy of the struck proton one can
neglect the motion of the spectator nucleons during the
propagation of the fast struck proton through the residual
nucleus.
The averaging over the spectator
nucleons positions of
the reduced nuclear matrix element squared which was obtained
through the wave equation relevant to the struck proton
at fixed $\tau$ is performed in Eq. (\ref{eq:2.8}).
It is clear that the averaging over $\tau$ is but the
evaluation of the diagonal matrix element of the subsystem
(the struck proton) in the momentum space.
In another words the difference between the
treatment of the FSI effects in the DWIA
approach and in the Glauber model can be formulated as a
difference in the order of operation. Schematically the order
of operation in the DWIA is as follows:
\begin{enumerate}
\item{Averaging over the spectator nucleon positions (the
evaluation of the effective optical potential).}
\item{Solving of the wave equation for the struck
proton wave function using the effective optical potential
and calculation of the reduced nuclear matrix element squared.}
\end{enumerate}
In the case of the Glauber model the reverse
order is used: \begin{enumerate}
\item{Solving of the wave equation for the struck proton
wave function at fixed positions of the spectator nucleons
and computing of the reduced nuclear matrix element squared.}
\item{Averaging over the spectator nucleon positions.}
\end{enumerate}

In contrast to the optical potential FSI factor (\ref{eq:3.2}) the
Glauber model factor (\ref{eq:2.27}) has a non-factorizable form due
to the presence of the term \GG. The interaction between the two
trajectories of the struck proton in the Glauber FSI factor
(\ref{eq:2.27}) connected with
\GGp term is a consequence of the averaging over
the spectator nucleon positions after computing the matrix element
squared for fixed the spectator configuration $\tau$.
The physical origin of the \GGp term in the FSI factor $\Phi$
(\ref{eq:2.27}) is the incoherent (see section 4)
rescatterings of the struck proton on the spectator nucleons
during its propagation through the target nucleus. Namely
the sum over the nucleus excitations created by the elastic
rescatterings of the ejected proton leads to the non-factorizable
expression (\ref{eq:2.27}).

There is a formal  analogy between the optical model FSI factor
(\ref{eq:3.2}) and the Glauber model factor, if the
\GGp term in exponent (\ref{eq:2.27}) is excluded.
Such a reduced attenuation factor,
$\Phi_{opt}^{Gl}(\vec{r}_{1},\vec{r}_{1}^{\,'})$, takes on a
factored form as a function of $\vec{r}_{1}$ and $\vec{r}_{1}^{\,'}$
\beq
\Phi_{opt}^{Gl}(\vec{r}_{1},\vec{r}_{1}^{\,'})=
S_{opt}^{Gl}(\vec{r}_{1}) S_{opt}^{Gl*}(\vec{r}_{1}^{\,'})\,,
\label{eq:3.3}
\eeq
with
\beq
S_{opt}^{Gl}(\vec{r})=\exp\left[-\frac{1}{2}\sigma_{tot}(pN)
(1-i\alpha_{pN})t(\vec{b},z)\right]\,.
\label{eq:3.4}
\eeq
The integral nuclear transparency (\ref{eq:2.16}) and the missing
momentum distribution (\ref{eq:2.20}) calculated with using FSI factor
(\ref{eq:2.27}) and (\ref{eq:3.3}) differ substantially. Our
numerical results give clear cut evidence that  the \GGp term in
(\ref{eq:2.27}) becomes very important in the
region $|\vec{p}_{m}|\gsim 200$ MeV/c. This is a consequence of the
short range (in the variable  $(\vec{r}_{1}-\vec{r}_{1}^{\,'})$)
"interaction" between two trajectories in the FSI factor
(\ref{eq:2.27}). It is clear that such a short range
"interaction" can not be modeled in the optical potential approach
even at the expense
of any modification of the attenuation factor (\ref{eq:3.2}).
Thus, the  DWIA, which was successful in the region of low $Q^{2}$,
can not be extended to description of inclusive $(e,e'p)$
reaction at high $Q^{2}$.
The peculiarity of the high-$Q^{2}$ region is that in this
case both the coherent rescattering of the struck proton on
the spectator nucleons and the incoherent ones practically
do not change the direction of the proton momentum. For this
reason they need to be treated on the same footing. On the
contrary, at low $Q^{2}$, when the energy of the struck proton
is small, every incoherent rescattering of the struck proton
leads to a considerable loss of the proton energy-momentum.
As a result, the flux of the outgoing proton plane wave is
suppressed.
This effect is modeled in the DWIA by the imaginary part of the
effective optical potential. Thus, the DWIA and the Glauber model
appear to be applicable for description of $(e,e'p)$ scattering in
different kinematical domains, at low $Q^{2}$ and at high $Q^{2}$,
respectively.

It is interesting that at high $Q^{2}$ the optical potential form of
the Glauber model FSI factor (\ref{eq:3.3}) still has a certain
applicability domain. Namely,
in a certain sense the attenuation factor (\ref{eq:3.3})
describes the FSI in exclusive $(e,e'p)$ scattering,
when only the events with one knocked out nucleon (proton)
are allowed. This connection between the FSI factor (\ref{eq:3.3}) and
exclusive $(e,e'p)$ reaction takes place in the independent particle
shell model. One can show that in the shell model
without $NN$  correlation the FSI factor (\ref{eq:3.3}) corresponds to
the situation when the sum over the residual nucleus states includes
only the states which arise from the ground state of the target
nucleus after removing one of the protons.
Indeed, in this case the index $f$ in Eq. (\ref{eq:2.1})
indicates one of the states of the target nucleus occupied
by the protons. Thus Eq. (\ref{eq:2.1}) takes the form
\beq
M_{f}=\int d^{3}\vec{r}_{1}...d^{3}\vec{r}_{A}
\Psi_{i,f}^{*}(\vec{r}_{2},...,\vec{r}_{A})
\Psi_{i}(\vec{r}_{1},...,\vec{r}_{A})
S(\vec{r}_{1},...\vec{r}_{A})\exp(i\vec{p}_{m}\vec{r}_{1})\,,
\label{eq:3.5}
\eeq
here, as in Eq. (\ref{eq:2.182}),  $\Psi_{i,f}$ is
the $(A-1)$-body wave function of the system of $(A-1)$
nucleons obtained after removing of the proton in the state $f$ from
the target nucleus. Making use of the Slater determinant form of
$\Psi_{i,f}$ and $\Psi_{i}$
one can obtain to the leading order in $1/A$ for (\ref{eq:3.5})
\beq
M_{f}=\frac{1}{\sqrt{Z}}
\int d^{3}\vec{r}_{1}...d^{3}\vec{r}_{A}
\exp(i\vec{p}_{m}\vec{r}_{1})
\phi_{f}(\vec{r}_{1})
\rho_{f}(\vec{r}_{2},...,\vec{r}_{A})
S(\vec{r}_{1},...\vec{r}_{A})\,,
\label{eq:3.6}
\eeq
here $\rho_{f}$ is $(A-1)$-body nuclear density defined by
Eq. (\ref{eq:2.182}). As in section 2 we will use a factored
representation (\ref{eq:2.185}) for this distribution. Then,
making use of Eqs.~(\ref{eq:2.2},\ref{eq:2.24})
we arrive at the following
expression for the matrix element (\ref{eq:3.5})
\beq
M_{f}=\frac{1}{\sqrt{Z}}
\int d^{3}\vec{r}_{1}\exp(i\vec{p}_{m}\vec{r}_{1})
\phi_{f}(\vec{r}_{1})S_{opt}^{Gl}(\vec{r}_{1})\,.
\label{eq:3.7}
\eeq
Taking the sum of matrix elements
(\ref{eq:3.7}) squared
over the states $f$,
immediately leads to the formula
(\ref{eq:2.20}) with the reduced FSI factor (\ref{eq:3.3}) instead
of the whole one (\ref{eq:2.27}). It is worth noting that from the
quantum mechanical point of view, the FSI factor (\ref{eq:3.3})
describes the FSI effects from coherent rescatterings of the
struck proton.

Making use of the FSI factors (\ref{eq:2.27}), (\ref{eq:3.3}) one can
obtain for
the integral nuclear transparency in quasielastic $(e,e'p)$ scattering
\beq
T_{A}^{\,inc}=\int
d^{2}\vec{b}dz\rho_{A}(\vec{b},z)\exp[-\sigma_{in}(pN)t(\vec{b},z)]\,,
\label{eq:3.8}
\eeq
($\sigma_{in}(pN)=\sigma_{tot}(pN)-\sigma_{el}(pN)$)
for the inclusive reaction, and
\beq
T_{A}^{\,exc}=\int
d^{2}\vec{b}dz\rho_{A}(\vec{b},z)\exp[-\sigma_{tot}(pN)t(\vec{b},z)]\,
, \label{eq:3.9}
\eeq
for the exclusive process.

Thus the integral nuclear transparency is controlled by
the inelastic proton-nucleon cross section in the inclusive case
and by the total proton-nucleon cross section in the exclusive
one. The physical origin of this difference is obvious.
The incoherent rescatterings of the struck proton in the
nuclear medium do not reduce the flux of the ejected proton. For
this reason the attenuation is controlled by the inelastic
proton-nucleon cross section in the inclusive case. In the exclusive
$(e,e'p)$ reaction the processes with knocking out one ( or more)
of the spectator
nucleons by the struck proton due to elastic incoherent rescatterings
are forbidden. As a result, the attenuation is controlled by
the total proton-nucleon cross section, and
$T_{A}^{\,exc}<T_{A}^{\,inc}$.

As we already emphasized the correspondence between the FSI
factor (\ref{eq:3.3}) and exclusive $(e,e'p)$ scattering takes place
in the idealized shell model.
Evidently, because of short range $NN$ correlations, the
above relationship between the missing momentum
distribution calculated with the FSI factor (\ref{eq:3.3}) and the
observed exclusive cross section will partly be lost.
Indeed, $(e,e'p)$ scattering on the proton of the
correlated $NN$ pair leads to an ejection of the spectator
nucleon of the correlated $NN$ pair \cite{NN1,NN2}, and
the corresponding final state will not fall into the exclusive
category. However, at small missing momenta $\vec{p}_{m}$ the
probability of ejection of spectators must be small, because
the typical momenta of nucleons in the correlated pair are of the same
order in magnitude and $\gsim k_{F}$, and triggering on small
$\vec{p}_{m}$ one effectively suppresses the contribution from
correlated $NN$ pairs.

\section{Incoherent rescatterings and transverse missing momentum
distribution}
Eqs. (\ref{eq:3.8}), (\ref{eq:3.9}) show that in the case of the
integral
nuclear transparency a simple quasiclassical treatment of the
FSI effects in $(e,e'p)$ scattering is possible to a certain extent.
At the same time it is clear, that the missing momentum distribution
(\ref{eq:2.20}) ( or $T_{A}(\vec{p}_{m})$  as a function of
$\vec{p}_{m}$) does not admit a treatment at the classical level.
The integration
over $\vec{r}_{1}$ and $\vec{r}_{1}^{\,'}$ in Eq. (\ref{eq:2.20})
shows that experimentally observed cross section of $(e,e'p)$
scattering at some $\vec{p}_{m}$ is a result of manifestly quantum
interference of amplitudes with different positions
where virtual photon strikes the proton. We would like to
emphasize  that in the case of the three-dimensional missing momentum
distribution (\ref{eq:2.20}) the FSI effects from the incoherent
rescatterings (connected
with the term \GG in the FSI factor (\ref{eq:2.27})) taken separately
also can not be treated
at a classical level. Indeed, the low limit in the $z$-integration for
the term \GG  in Eq. (\ref{eq:2.27}) is equal to the maximum value
of the low limits of the $z$-integration in the terms $\propto\Gamma$
and $\propto\Gamma^{*}$. It implies that there is a part of the
struck proton's trajectory where the incoherent rescatterings
are forbidden. Due to this fact a probabilistic interpretation
of the effects connected with the \GGp term in the FSI factor
(\ref{eq:2.27})
in the case of the nonintegrated distribution (\ref{eq:2.20}) is not
possible.

In the case of the FSI factor (\ref{eq:2.31}) related to the $p_{m,z}$
integrated $\vec{p}_{m\perp}$ distribution (\ref{eq:2.29}), the low
limits of $z$-integrations in the terms
\G and \GG are equal. It enables one to extend, to a certain extent,
the probabilistic treatment, that is  possible for the
integral nuclear transparency (\ref{eq:3.8}), to the transverse
missing momentum distribution (\ref{eq:2.29}). To demonstrate this
fact it is convenient to rewrite (\ref{eq:2.29}) in the following
form
\bea
w_{\perp}(\vec{p}_{m\perp})=\frac{1}{(2\pi)^{2}}
\int d^{2}\vec{b}dz d^{2}\vec{\Delta}
\exp(i\vec{p}_{m\perp}\vec{\Delta})
\rho(\vec{b}+\frac{1}{2}\vec{\Delta},z,
\vec{b}-\frac{1}{2}\vec{\Delta},z)\nonumber\\  \times
S_{opt}^{Gl}(\vec{b}+\frac{1}{2}\vec{\Delta},z)
S_{opt}^{Gl*}(\vec{b}-\frac{1}{2}\vec{\Delta},z)
\exp\left[\eta(\vec{\Delta})\sigma_{el}(pN)t(\vec{b},z)\right]\,.
\label{eq:4.1}
\eea
Let us introduce a local momentum distribution including
the distortion effects at the level of the optical form of the
Glauber FSI factor (\ref{eq:3.3}) defined as follows
\bea
w_{\perp,opt}(\vec{b},z,\vec{p}_{m\perp})=\frac{1}{(2\pi)^{2}}
\int d^{2}\vec{\Delta}
\exp(i\vec{p}_{m\perp}\vec{\Delta})
\rho(\vec{b}+\frac{1}{2}\vec{\Delta},z,
\vec{b}-\frac{1}{2}\vec{\Delta},z) \nonumber \\ \times
\rho_{A}^{-1}(\vec{b},z)
S^{Gl}_{opt}(\vec{b}+\frac{1}{2}\vec{\Delta},z)
S_{opt}^{Gl*}(\vec{b}-\frac{1}{2}\vec{\Delta},z)\,.
\label{eq:4.2}
\eea
The local distribution (\ref{eq:4.2}) is normalized as
$$
\int d^{2}\vec{p}_{m\perp}
w_{\perp,opt}(\vec{b},z,\vec{p}_{m\perp})=
|S_{opt}^{Gl}(\vec{b},z)|^{2}\,.
$$

Making use of the local missing momentum distribution (\ref{eq:4.2})
after expansion of the last exponential factor in Eq. (\ref{eq:4.1})
in a power-series, one can represent the formula (\ref{eq:4.1}) in a
form of $\nu$-fold rescattering series
\beq
w_{\perp}(\vec{p}_{m\perp})=\sum\limits_{\nu=0}^{\infty}
w_{\perp}^{\nu}(\vec{p}_{m\perp})\,,
\label{eq:4.4}
\eeq
where the zeroth order term is given by
\beq
w_{\perp}^{0}(\vec{p}_{m\perp})=\int d^{2}\vec{b}dz
\rho_{A}(\vec{b},z) w_{\perp,opt}(\vec{b},z,\vec{p}_{m\perp})\,,
\label{eq:4.5}
\eeq
and the contribution of $\nu$-fold component for $\nu\geq 1$
reads
\bea
w_{\perp}^{\nu}(\vec{p}_{m\perp})=\frac{1}{\nu !}\int d^{2}\vec{b}dz
\rho_{A}(\vec{b},z) t^{\nu}(\vec{b},z)\nonumber\\ \times
\int\prod\limits_{i=1}^{\nu} d^{2}\vec{q}_{i}
\left(\frac{1}{\pi}\frac{d\sigma_{el}(pN)}{dq^{2}_{i}}\right)
w_{\perp,opt}(\vec{b},z,\vec{p}_{m\perp}-
\sum\limits_{j=1}^{\nu}\vec{q}_{j})\,.
\label{eq:4.6}
\eea
Eqs. (\ref{eq:4.4})-(\ref{eq:4.6}) embody the representation of the
transverse missing momentum distribution in a form when
all the quantum distortion effects are contained in the
local missing momentum distribution computed with the FSI factor
without \GGp term. The contribution from the incoherent rescatterings
connected with \GGp term admits a probabilistic reinterpretation.

The transverse missing momentum distribution (\ref{eq:2.29})
obtained using the closure relation (\ref{eq:2.7}) is
appropriate
to inclusive $(e,e'p)$ scattering. Nevertheless the representation
(\ref{eq:4.4}) can be used to estimate the contribution to the cross
section
of this process from the events with fixed number of the knocked out
(recoil) nucleons. Our numerical results show that dominant
contribution
to the transverse missing momentum distribution in the region
$p_{\perp}\lsim k_{F}$ comes from the terms with $\nu\leq 1$. It means
that inclusive cross section of $(e,e'p)$ scattering in the above
kinematical domain is saturated by the events
without and with one knocked out nucleon (besides the ejected proton).
At high $p_{m\perp}$ the contribution from the terms with
$\nu > 1$ becomes also important. The role of the incoherent
rescatterings in the region $p_{\perp}\gsim k_{F}$ was
recently discussed in \cite{Jresc}.
It was shown, in
particular,
that incoherent rescatterings lead to a large tail in the missing
energy distribution. The last exponent in Eq. (\ref{eq:4.1})
has the most steep dependence on the variable $\vec{\Delta}$.
As a result, at high $p_{\perp}$ the incoherent
rescattering effects become more important than the distortion
effects. In \cite{Jresc} only incoherent rescatterings were
taken into account.
As we will see in the region $p_{\perp}\lsim k_{F}$
we are interested in the present paper,
the relative magnitude of the FSI effects connected with
distortion effects and incoherent rescatterings are of the
same order and both of them  must be taken into account
simultaneously.

One remark on the physical interpretation of the
incoherent rescatterings in quasielastic $(e,e'p)$ reaction,
is in order. We would like to emphasize a somewhat formal nature of
extension  (\ref{eq:4.4}). It
does not mean that the FSI effects connected with
incoherent rescatterings allow the classical treatment.
For instance, as we will see below, the representation (\ref{eq:4.4})
even does not imply
that the momentum transfers in the incoherent rescatterings of the
struck proton on spectator nucleons are purely transverse.
For this reason, in particular, it is not possible to model
the FSI effects associated with the incoherent rescatterings of
the struck proton in the nuclear medium by virtue of the
Monte-Carlo approach.

\section{Longitudinal missing momentum distribution and
applicability limits of Glauber model}
The failure of the quasiclassical probabilistic treatment of the
incoherent rescatterings becomes especially evident in the
case of the longitudinal missing momentum distribution. Indeed,
naively, from the classical point of view, one can expect that this
distribution is not affected by the elastic rescatterings of the
struck proton on the spectator nucleons. In fact, as one can see
from Eqs. (\ref{eq:2.20}), (\ref{eq:2.27}), because of the last
term in the exponent in Eq. (\ref{eq:2.27}),
the incoherent rescatterings must affect, and have a quite
nontrivial impact on, the longitudinal missing momentum distribution.

It is worth noting that even from a simple qualitative
quantum mechanical consideration one can understand that the
incoherent rescattering must influence upon the longitudinal
missing momentum distribution. Indeed, let $\Delta l$ be the
distance between the point where the virtual photon strikes
the ejected proton and the point where incoherent scattering off the
spectator nucleon takes place. Then, it is evident from the
uncertainty relation that the momentum transfer in the
$pN$ scattering has uncertainty $\Delta k \sim 1/\Delta l$.
In the case of sufficiently small $\Delta l$ the longitudinal
momentum transfer can be comparable to the  transverse one.
Indeed it is evident from the uncertainty relation that in the
situation when $\Delta l$  is about the proton size
both longitudinal and transverse momentum transfer will be
about the inverse proton radius.
Thus, one can conclude that in the region of high $|p_{m,z}|$,
$p_{m\perp}$
($|p_{m,z}|\sim p_{m\perp}$)  the incoherent rescatterings of
the struck proton on the adjacent spectator nucleons must
considerably affect the missing momentum distribution
as compared to the PWIA case.
As we will see the numerical results actually show that
the additional \GGp term in the FSI factors (\ref{eq:2.27}),
(\ref{eq:2.30}) gives rise a considerable tail at high
$|p_{m,z}|$, which is missed if the optical potential
form (\ref{eq:3.3}) of the FSI factor is used.

Formally, the sensitivity of the longitudinal missing momentum
distribution to the incoherent rescatterings is connected with
the above mentioned peculiarity in the $z$-integration for the
term \GG in Eq. (\ref{eq:2.27}). Indeed,
 the function $\Phi_{opt}^{Gl}(\vec{r}_{1},\vec{r}_{1}^{\,'})$
defined by Eq. (\ref{eq:3.3}) has a smooth behavior in the variable
$\xi=z_{1}-z_{1}^{\,'}$
at the point $\xi=0$. On the contrary, the FSI factor
(\ref{eq:2.27})
as a function of $\xi$ has a discontinuous derivative
with respect to $\xi$ at the point $\xi=0$. The origin of
this discontinuity is the appearance in the \GGp term of
the nonanalytical function $max(z_{1},z_{1}^{'})$ as the low limit
in the $z$-integration. Evidently,
the singular behavior of the integrand in Eq. (\ref{eq:2.20}) ( and
Eq. (\ref{eq:2.28}) as well) in the variable $\xi$, after Fourier
transform will show itself up as an anomalous behavior of the missing
momentum distribution at high longitudinal momenta.

      Let us proceed with the analysis of the situation
with the longitudinal momentum distribution for the case
of the $\vec{p}_{m\perp}$ integrated distribution (\ref{eq:2.28}).
The absence of the integration over $(\vec{b}-\vec{b}^{\,'})$
makes this case considerably more simple, as compared to the
nonintegrated one (\ref{eq:2.20}), for a qualitative analysis.

It is convenient to rewrite the longitudinal FSI factor
(\ref{eq:2.30}) in such a form
\beq
\Phi_{z}(\vec{b},z,z^{'})=
\Phi_{z,opt}^{in}(\vec{b},z,z^{'})
C_{1}(\vec{b},z,z^{'})C_{2}(\vec{b},z,z^{'})\,,
\label{eq:5.1}
\eeq
with
\beq
\Phi_{z,opt}^{in}(\vec{b},z,z^{'})=\exp\left[-\frac{1}{2}
\sigma_{in}(pN)t(b,z)-\frac{1}{2}\sigma_{in}(pN)t(b,z^{'})\,,
\right]
\label{eq:5.2}
\eeq
\beq
C_{1}(\vec{b},z,z^{'})=\exp\left[\frac{i}{2}
\sigma_{tot}(pN)\alpha_{pN}\left(t(b,z)-t(b,z^{'})\right)\right]\,,
\label{eq:5.3}
\eeq
\beq
C_{2}(\vec{b},z,z^{'})=\exp\left[-\frac{1}{2}
\sigma_{el}(pN)\left|t(b,z)-t(b,z^{'})\right|\right]\,.
\label{eq:5.4}
\eeq

In Eq. (\ref{eq:5.1}) we singled out from the FSI factor
(\ref{eq:2.28}) the function
$\Phi_{z,opt}^{in}(\vec{b},z,z^{'})$. In a certain sense it can be
interpreted
as the optical potential FSI factor taking into account only
the distortion of the plane wave connected with the real inelastic
interactions of the struck proton in the nuclear medium. We will refer
to the corresponding missing momentum distribution as
$w_{z,opt}^{in}(p_{m,z})$.
$\Phi_{z,opt}^{in}(\vec{b},z,z^{'})$ is a symmetric function of
$z$, $z^{'}$. For this reason $w_{z,opt}^{in}(p_{m,z})$ is an even
function of $ p_{m,z}$.

 For the purpose of the qualitative analysis
in the case of $A\gg1$ one can approximate the functions
$C_{1,2}(\vec{b},z,z^{'})$ by the following expressions
\beq
C_{1}(\vec{b},z,z^{'})=\exp\left[-ik_{1}
(z-z^{'})\right]\,,
\label{eq:5.5}
\eeq
\beq
C_{2}(\vec{b},z,z^{'})=\exp\left[-k_{2}
|z-z^{'}|\right]\,,
\label{eq:5.6}
\eeq
where
\beq
k_{1}=\frac{1}{2}
\sigma_{tot}(pN)\alpha_{pN}\langle n_{A}\rangle\,,
\label{eq:5.7}
\eeq
\beq
k_{2}=\frac{1}{2}
\sigma_{el}(pN)\langle n_{A}\rangle\,,
\label{eq:5.8}
\eeq
and $\langle n_{A}\rangle$ is the average nuclear
density.

Then, using these approximations, the longitudinal missing momentum
distribution may be represented in such a convolution form
\beq
w_{z}(p_{m,z})\approx\frac{1}{2\pi}\int dk
w_{z,opt}^{in}(p_{m,z}-k_{1}-k)c_{2}(k)\,.
\label{eq:5.9}
\eeq
Here we used notation $c_{2}(k)$ for the Fourier transform
of the factor $C_{2}$ approximated by formula (\ref{eq:5.6}).
It is clear from Eq. (\ref{eq:5.9})  that the major effect of the
nonzero
real part of the $pN$-amplitude contained in the factor $C_{1}$
is a shift of the longitudinal missing momentum distribution by
$k_{1}$ \cite{JNE18}. In the region $Q^{2}\sim 2-10$ GeV$^{2}$
the shift is quite large $k_{1}\sim 20$ MeV/c.
Thus nonzero $\alpha_{pN}$ leads to asymmetry of $p_{m,z}$
distribution about $p_{m,z}=0$.
The role of the factor $C_{2}$ is
more interesting. The Fourier transform of the factor $C_{2}$
for the case of the approximation (\ref{eq:5.6}) is given by
\beq
c_{2}(k)=\int d\xi \exp(ik\xi)\exp(-k_{2}|\xi|)
=\frac{2k_{2}}
{k^{2}+k_{2}^{2}}\,.
\label{eq:5.10}
\eeq

 In the kinematical domain  we are interested in
$k_{2}\sim 10-20$ MeV/c, and the inequality $k_{2}\ll k_{F}$ is
satisfied. It means that the Fourier transform of $C_{2}$ when
approximation (\ref{eq:5.6}) is used has a
form of a sharp peak with width that is much less then the width of
the Fermi distribution.
In the real situation when exact expression (\ref{eq:5.4}) is used
the width of the peak of the Fourier transform of the function
$C_{2}(\vec{b},z,z^{'})$ in the variable $\xi=(z-z^{'})$
will be controlled by  the inverse nucleus size because
$k_{2}\lsim 1/R_{A}$ for real nucleus. Due to
inequality $k_{F}\gg 1/R_{A}$ this width again turns out to be
much less than the width of the Fermi distribution.
The finite nucleus size will not change the asymptotic
law $c_{2}(k)\propto k^{-2}$ connected only with a nonanalytical
behavior of expressions (\ref{eq:5.4}), (\ref{eq:5.6}) at the point
$z=z^{'}$. From the above consideration it is clear that
at $|p_{m,z}|\ll k_{F}$ the factor $c_{2}(k)$ in the convolution
representation (\ref{eq:5.9}) acts like $\delta$-function. Hence, at
small $|p_{m,z}|$ we will have
\beq
w_{z}(p_{m,z})\approx w_{z,opt}^{in}(p_{m,z}-k_{1})\,.
\label{eq:5.11}
\eeq
 However, at sufficiently large
$|p_{m,z}|\gsim k_{F}$ the $p_{m,z}$ dependence of $w_{z}(p_{m,z})$
will be controlled by the asymptotic behavior of the factor
$c_{2}(k)$, and the following regime will set in
\beq
w_{z}(p_{m,z})\propto p_{m,z}^{-2}\,.
\label{eq:5.12}
\eeq

Our numerical results show that already in the region
$|p_{m,z}|\sim 200-300$ Mev/c there is a considerable
deviation from the approximate formula (\ref{eq:5.11}). The missing
momenta at which the onset of the regime (\ref{eq:5.12}) takes place
are considerably greater than $k_{F}$. The independent particle
shell model used in the present paper is not applicable for
analysis of $(e,e'p)$ scattering in this kinematical domain.

Let us consider in detail the physical reason for the appearance of
the longitudinal momentum transfer in the incoherent rescatterings
of the struck proton in $(e,e'p)$ reaction. The origin of the
longitudinal momentum transfer in $(e,e'p)$ scattering is connected
with the absence in this case of an incoming proton plane wave.
Indeed, as it was mentioned above the missing momentum
distribution (\ref{eq:2.20}) from the
quantum mechanical point of view corresponds to the interference of
the amplitudes with different positions at which the virtual photon
strikes the proton. For each of this amplitudes the wave
function of the spectator nucleons, after the struck proton
leaves the target nucleus, will be distorted along
the straight line that begins from the point where the photon-proton
interaction takes place. It is clear that decomposition of the
distortion of the spectator nucleon wave function into the plane
waves contains besides the components with transverse momentum
the components with longitudinal momentum. The asymptotic
behavior (\ref{eq:5.12}) of the longitudinal momentum distribution
is a consequence of the discontinuous distortion of
the spectator nucleon wave functions. This discontinuity is connected
with $\theta$-function appearing in the Glauber model attenuation
factor (\ref{eq:2.2}). It is evident that allowance for
the finite longitudinal size, $d_{int}$, of the region
where proton-nucleon
interaction takes place, must lead to a smearing of the
sharp edge of the $\theta$-function in Eq. (\ref{eq:2.2}). Evidently,
$d_{int}$ is about the proton radius. As a consequence of the
smearing of $\theta$-function in Eq. (\ref{eq:2.2}) the
$p_{m,z}^{-2}$ law (\ref{eq:5.12}) will be replaced by a somewhat
steeper decrease at high $|p_{m,z}|$.  We can suggest
as a generalization of Eq. (\ref{eq:2.2}) to the case of finite
$d_{int}$ the same equation in which one of the following replacements
is made
\beq
\theta(z) \Rightarrow \frac{\sqrt{\pi}}{d_{int}}
\int \limits_{z}^{\infty} d\xi \exp(-\xi^{2}/d_{int}^{2})\,,
\label{eq:5.13}
\eeq
or

\beq
\theta(z) \Rightarrow \frac{1}{2}
\left[1+\tanh(z/d_{int})\right]\,.
\label{eq:5.14}
\eeq
Of course, there are no serious theoretical
motivations to use either of the  prescriptions (\ref{eq:5.13}) or
(\ref{eq:5.14}) at such $|p_{m,z}|$,
where the missing momentum distributions corresponding to the
attenuation factor (\ref{eq:2.2}) and obtained with using
replacement
(\ref{eq:5.13}) or (\ref{eq:5.14}) differ strongly.
Nevertheless,
they can be used to clarify the applicability limits of the standard
Glauber model. Fortunately, it turns out that the kinematical range
of the longitudinal missing momentum, where the Glauber model is
still applicable, is quite broad. In \cite{He4,D2} it was checked
that $\theta$-function Ansatz in the Glauber attenuation factor
in the case of deuteron works very well in the region
$|p_{m,z}|\lsim 500$ MeV/c. Our numerical calculations in the
kinematical domain  $|p_{m,z}|\lsim 300$ MeV/c also show that
introduction of the finite interaction size $d_{int}\sim 1$ $fm$
practically does not change the standard Glauber model predictions
obtained with using $\theta$-function in Eq. (\ref{eq:2.2}).

In connection with above discussion it is worth noting the following.
In our analysis we neglected the short range $NN$ correlations.
Due to $NN$ repulsive core the probability to find in the target
nucleus two nucleons at the same point really is suppressed.
It means that at high  $|p_{m,z}|$ even without allowance for
finite $d_{int}$ the tail longitudinal missing momentum distribution
will decrease considerably steeper when the short range $NN$
correlation are taken into account. Since  $NN$ correlation
radius $\sim 1$ $fm$ this effect can be neglected in the
region $|p_{m,z}|\lsim 300$ MeV/c likewise one from the finite
$d_{int}$.

The found incompleteness of the Glauber model in the case
of $(e,e'p)$ scattering in high-$|p_{m,z}|$ region makes
questionable the possibility of using the measured
missing momentum distribution for obtaining the information on the
short range $NN$ correlations in nuclei. It is clear now
that besides the short range $NN$ correlations,
the missing momentum distribution  at high $|p_{m,z}|$
probes the nucleon structure as well.
Notice, that the sensitivity of high-$|p_{m,z}|$ tails to the
nucleon size survives at high $Q^{2}$ as well. For this reason we will
face the same problem in the analysis of $(e,e'p)$ scattering at high
$Q^{2}$, in the CT regime of large contribution from the off-diagonal
inelastic rescatterings.

In conclusions of this section we would like to emphasize
that in the case of reaction $(e,e'p)$ we face a situation which
is quite different from the small angle hadron-nucleus scattering.
In the latter case the incoming hadron plane wave exists. Hence, the
$\theta$-function effect in the Glauber model attenuation factor
(\ref{eq:2.2}) disappears. As a result, the Glauber model predictions
do not contain uncertainties connected with the finite longitudinal
size $d_{int}$. However, it is clear that this problem will arise in
the case of quasielastic $(p,2p)$ scattering at large angle.
The analysis of the reaction $(p,2p)$ in the kinematical range
of the BNL experiment \cite{BNLexp} will be presented elsewhere.
\section{Numerical results}
In this section we present our numerical results for the missing
momentum distribution and nuclear transparency in quasielastic
$(e,e'p)$ scattering obtained in the Glauber formalism.
To avoid complications with the target nucleus spin we restricted
ourselves to the case of closed shell nuclei $^ {16}O$ and $^{40}Ca$.
We adjusted the oscillator shell model frequency, $\omega_{osc}$, for
these two nuclei to reproduce the experimental value of the
root-mean-square radius of the charge distribution,
$\langle r^{2}\rangle^{1/2}$. We used the values \cite{Atdata}
$\langle r^{2}\rangle^{1/2}=2.73$ $fm$ for $^{16}O$, and
$\langle r^{2}\rangle^{1/2}=3.47$ $fm$ for $^{40}Ca$,
which correspond to the oscillator
radius, $r_{osc}=(m_{p}\omega_{osc})^{-1/2}$, equal to
1.74 $fm$ for $^ {16}O$ and  1.95 $fm$ for $^{40}Ca$.
The difference between the charge distribution and the proton
nuclear density connected with the proton charge radius was
taken into account.

As it was stated in section 2 we use the exponential parameterization
of the proton-nucleon elastic amplitude. The diffraction slope of the
$pN$ scattering was estimated from the relation
\beq
B_{pN}\approx \frac{\sigma_{tot}^{2}(pN)(1+\alpha_{pN}^{2})}
{16\pi \sigma_{el}(pN)}\,.
\label{eq:6.1}
\eeq
In our calculations we define the $pN$ cross sections and
$\alpha_{pN}$ as mean values of these quantities for
$pp$ and $pn$ scattering. We borrowed the experimental data
on $pp$, $pn$ cross sections and $\alpha_{pp}$, $\alpha_{pn}$ from
the recent review \cite{Lehar}. From the point of view of the $Q^{2}$
dependence of the Glauber model predictions for the missing momentum
distribution in $(e,e'p)$ scattering in the region
$Q^{2}\sim 2-10$ GeV$^{2}$, especially important is the energy
dependence of $B_{pN}$ and $\sigma_{el}(pN)$. We remind that
the typical kinetic energy
of the struck proton $T_{kin}\approx Q^{2}/2m_{p}$.
Typically $B_{pN}$ rises from $B_{pN}\approx 4.5$ GeV$^{-2}$ at
$Q^{2}=2$ GeV$^{2}$ to  $B_{pN}\approx 8$ GeV$^{-2}$ at
$Q^{2}=10$ GeV$^{2}$, and $\sigma_{el}(pN)$ falls from
$\sigma_{el}(pN)\approx 23$ $mb$ at $Q^{2}=2$ GeV$^{2}$ to
$\sigma_{el}(pN)\approx 11.5$ $mb$ at $Q^{2}=10$ GeV$^{2}$.
In our kinematical domain $\sigma_{tot}(pN)$ slightly
decreases with $Q^{2}$, $\sigma_{tot}(pN)\approx 43.5\,mb$
at $Q^{2}=2$ GeV$^{2}$ and $\sigma_{tot}(pN)\approx 40\,mb$
at $Q^{2}=10$ GeV$^{2}$.
We remind that in the Glauber model the trajectories of the
high energy particles are assumed to be straight lines.
In our case the struck proton momentum $\sim 2$ GeV at $Q^{2}\sim 2$
GeV$^{2}$.
In this region of momenta the mean value of the momentum transfer in
$pN$ scattering is $\sim 1/\sqrt{B_{pN}}\sim 0.45$ GeV/c. Thus
there are reasons to believe that the Glauber formalism is still
reliable at lower bound of the kinematical domain we are
interested in the present paper $Q^{2}\sim 2-10$ GeV $^{2}$.

To illustrate the role of the incoherent rescatterings in
process $(e,e'p)$  we present a systematical comparison of
results obtained for the full Glauber theory with
\GGp term in the FSI factor (\ref{eq:2.27}) included, and the
truncated version when \GGp term not
included. We remind that from the point of view of
the independent particle shell model these two versions are relevant
to the inclusive and exclusive conditions in $(e,e'p)$
scattering, respectively.

The results for the integral nuclear transparencies $T_{A}^{inc}$
and $T_{A}^{exc}$ defined by Eqs. (\ref{eq:3.8}), (\ref{eq:3.9})
are shown in Fig.~1. Because of the rise of $\sigma_{in}(pN)$,~
$T_{A}^{inc}$ slowly decreases in our kinematical range.
$T_{A}^{exc}$,
which is controlled by $\sigma_{tot}(pN)$, is approximately flat. As
one can
see from Fig.~1 the replacement of $\sigma_{in}(pN)$ by
$\sigma_{tot}(pN)$
considerably reduces the integral nuclear transparency.  We
use the computed
values of  $T_{A}^{inc}$   and $T_{A}^{exc}$ to obtain normalized to
unity missing momentum distribution (\ref{eq:2.17}) for the cases
with \GGp term in the FSI factor (inclusive $(e,e'p)$ scattering)
and without the one (exclusive $(e,e'p)$ scattering).

The integral nuclear transparencies for the case when the
kinematical domain $D$ in the definition (\ref{eq:1.1}) includes all
missing
momenta, depends only on the diagonal component of the one-body
nuclear
density matrix , $i.e.$, the nuclear density. On the contrary, the
missing momentum distribution and $T_{A}(\vec{p}_{m})$ as a functions
of $\vec{p}_{m}$ or the nuclear transparency for a certain
kinematical region $D$ , $T_{A}(D)$, defined by Eq. (\ref{eq:1.1})
are controlled by the
whole one-body density matrix. The major part of the results presented
in this section have been obtained using oscillator shell
model density matrix (\ref{eq:2.19}). In order to check the
sensitivity
of the results to the form of the one-body density matrix, we also
performed in a few cases the calculations using the LDA
parameterization (\ref{eq:2.32}) of the one-body density matrix.
A comparison of the results obtained in these two versions
is very interesting from the point of view of
clarifying the accuracy and applicability limits of the LDA
parameterization
(\ref{eq:2.32}) which is widely used in the literature for nuclei
with large nuclear mass number $A$.

In Fig.~2,~3 we show the angular dependence of the ratio
of the normalized missing momentum distribution
$n_{eff}(p_{m},\theta)$ to the Fermi momentum distribution
$n_{F}(p_{m})$ for
$p_{m}=$150, 200, 250 and  300 MeV/c at $Q^{2}=$2 and 10 GeV$^{2}$ .
The forward-backward  asymmetry of this ratio is  a consequence of the
nonzero real part of the elastic $pN$ amplitude.
 The appearance of a bump for $p_{m}=300$ MeV/c at $\theta\approx
100^{o}$ for $Q^{2}=2$ GeV$^{2}$ in the version with \GGp term
is connected with the fact that momentum transfers in the
incoherent rescatterings are predominantly transverse.
At $Q^{2}=10$ GeV$^{2}$, the bump evolved into the shoulder, which is
related to
the higher value of the diffraction slope. In contrast to the version
with \GGp term, in the case without \GGp term we obtained
a dip at $\theta\sim 80^{o}$ for $p_{m}\sim 250-300$
MeV/c. Thus we see that quantum FSI effects related to the
elastic rescatterings of the struck proton without the excitation
of the residual nucleus lead to a considerable distortion
of the outgoing proton plane wave.

Fig.~4 illustrates  $p_{m}$ dependence of the nonintegrated
nuclear transparence $T_{A}(p_{m},\theta)$ for $\theta=0^{o},\,
90^{o},\,180^{o}$ at $Q^{2}=2$ GeV$^{2}$. The results presented in
Fig.~4
more clearly demonstrate the relative role played by the absorption
effects connected with the terms \G and the incoherent
rescatterings effects related to the \GGp term in the full FSI factor.
We see that absorption leads  to appearance of a deep dip
in the nuclear transparency in $(e,e'p)$ reaction for exclusive
conditions at $p_{m}\sim 270-300$ MeV/c in the case of
transverse kinematics ($\theta=90^{o}$). With allowance for
the incoherent rescatterings (\GGp term is included) the
nuclear transparency in transverse kinematics  steeply
rises at $p_{m}\gsim 250$ Mev/c. Even in the parallel kinematics
($\theta=0^{o}$ and $\theta=180^{o}$) the effect of
\GGp term becomes
significant at $p_{m}\gsim 250$ Mev/c. This effect is a
manifestation of the longitudinal momentum transfer discussed
in section 5.
Thus the results presented in Fig.~4 show that the contribution
from the incoherent rescatterings becomes important only at
sufficiently large missing momenta, $|\vec{p}_{m}|\gsim 200-250$
MeV/c.

Notice, that Fig.~4 already demonstrates that  the
three-dimensional missing momentum
distribution has a substantially  non-factorizable dependence on the
transverse and longitudinal components of the missing momentum.
In order to demonstrate the degree of violation of the
$p_{m\perp}\!-\!p_{m,z}$
factorization we present in Fig.~5 the nuclear transparency
versus $p_{m\perp}$ at different
$p_{m,z}$ for $^{40}Ca$ at $Q^{2}=2$ GeV$^{2}$.
As one can see the  $p_{m\perp}\!-\!p_{m,z}$ factorization is
manifestly violated. Both versions, with \GGp term and
without one, demonstrate that $p_{m\perp}$ dependence of
$T_{A}(\vec{p}_{\perp},p_{m,z})$ is strikingly different from
that of the Fermi momentum distribution.
It makes it clear, in particular, that approach proposed in
ref.\cite{RJ} can not be justified.  As it was
explained in section 2, the approximation (\ref{eq:2.34}) of
ref. \cite{RJ} leads
to the almost $p_{m\perp}\!-\!p_{m,z}$ factorizable form of
$w(\vec{p}_{m})$
with the same $p_{m\perp}$ dependence as for the Fermi
momentum distribution.

In Fig.~6,~7 we show the behavior of the nuclear transparency
versus  $p_{m\perp}$ in the case when the kinematical domain $D$ in
the definition (\ref{eq:1.1}) includes all the longitudinal momenta.
{}From the point of view of the representation of
$w_{\perp}(\vec{p}_{m\perp})$ by the multiple scattering
series (\ref{eq:4.4}), the version with \GGp term kept corresponds
to the
situation when all incoherent rescatterings are taken into account.
The case without \GGp term in the FSI factor is equivalent to
keeping only the zeroth order term in series (\ref{eq:4.4}).
Besides these two cases we present in Fig.~6,~7 the results
for the case when two first terms in the series (\ref{eq:4.4}) are
included, $\nu=0$ and $\nu=1$.
As one can see from
Fig.~6,~7 in the kinematical domain under consideration the
mechanisms with $\nu=0,~1$ practically saturate the cross section of
inclusive $(e,e'p)$ reaction. The incoherent rescatterings of
the struck proton become important at $p_{m\perp}\gsim 200$ MeV/c.

The $p_{m,z}$ dependence of the nuclear transparency for
the case when the events with all transverse missing momentum
are included are shown in Fig.~8, the solid and dashed curves are
for the version with and without \GGp term. As one can see,
with increasing of nuclear
mass number, the development of the clear cut two-dip structure
takes place for the case without \GGp term. Thus the attenuation
effects connected with the FSI of the struck proton considerably
affect the missing momentum distribution as compared to the
PWIA case. Fig.~8 shows that the including of the \GGp term leads to
the appearance of large tails in the longitudinal missing momentum
distribution at $|p_{m,z}|\gsim 250$ MeV/c.
As in the case
of a purely parallel kinematics (see below), there is a considerable
asymmetry about $p_{m,z}=0$ connected with nonzero $\alpha_{pN}$.
The integral asymmetry, $A_{z}$, defined as
\beq
A_{z}=\frac{N(p_{m,z}>0)-N(p_{m,z}<0)}{N(p_{m,z}>0)+N(p_{m,z}<0)}
\label{eq:6.2}
\eeq
($N(p_{m,z}>0)$ and $N(p_{m,z}<0)$ are the number of events with
$p_{m,z}>0$  and $p_{m,z}<0$, respectively), is large,
$A_{z}\approx -(0.07-0.08)$, and approximately constant,
in our kinematical region ($Q^{2}\sim 2-10$ GeV$^{2}$).

Notice that the large $p_{m,z}$ asymmetry, obtained in the Glauber
model, obscures the study of the CT effects
in quasielastic $(e,e'p)$ scattering at $Q^{2}\lsim
10$ GeV$^{2}$ using
the experimental data on the dependence of the nuclear transparency on
the Bjorken $x$ (remind that $x\approx(1+p_{m,z}/m_{p})$)
\cite{JK,Jasym,Boffi1} . The point is that at $Q^{2}\sim
5-10$ GeV$^{2}$ the expected values of the asymmetry because of the
contribution of the inelastic (off-diagonal) scattering of the
struck proton on spectator nucleons \cite{Jasym,Boffi1} turn out to
be by factor 2-4 smaller than the above Glauber theory prediction
for $A_{z}$.
 The nonzero $\alpha_{pN}$ leading to the $p_{m,z}$
asymmetry in the Glauber model predictions is connected with
the contributions to the elastic $pN$ amplitude from the secondary
reggeons. As it is known \cite{reggeon} the reggeon exchange
requires a finite formation time increasing with hadron energy.
Thus one can expect that due to this effect real part of
the effective elastic $pN$ amplitude in the nuclear medium in
the case of $(e,e'p)$ scattering will partially differ from the one
measured in $pN$ scattering in vacuum. In the absence of
a rigorous theoretical model for the reggeon exchanges it will be
difficult to disentangle the CT contribution to the $p_{m,z}$
asymmetry. Nevertheless any rise of
$A_{z}$ with $Q^{2}$ will signal the onset of the CT effects,
since the finite formation time effects for the reggeon exchange
can only reduce $A_{z}$ predicted in the standard Glauber model.

It is instructive to compare
the $p_{m,z}$ dependence of the nuclear
transparency for the case of the parallel kinematics
, $p_{m\perp}=0$, $\theta=0^{o},~180^{o}$ shown in Fig.~4  with the
$p_{m,z}$ dependence
when all $\vec{p}_{m\perp}$ are included, Fig.~8. Such a comparison
very  clearly demonstrates, that the major contribution from
incoherent rescatterings into large-$|p_{m,z}|$ tails comes from
the region of the sufficiently large $p_{m\perp}$. The same
conclusion can be made from the curves presented in Fig.~5.
We remind that namely this pattern of the contribution
to the missing momentum distribution from the incoherent
rescatterings of the struck proton on the adjacent spectator nucleons
at high-$|p_{m,z}|$ region was predicted in section 5 on
the basis of the uncertainty relation. Thus, our numerical results
give clear cut evidence in favor of necessity to treat the
incoherent rescatterings of the struck proton in quasielastic
$(e,e'p)$ reaction in a quantum mechanical manner.

The above discussed peculiarities of the contribution of
the incoherent rescatterings into three-dimensional missing
momentum distribution in $(e,e'p)$ scattering are very
important from the point of view a comparison of the
theoretical predictions with experimental data on the
nuclear transparency obtained for a certain kinematical
domain $D$. Indeed, the integral nuclear transparencies
$T_{A}^{inc}$   and $T_{A}^{exc}$ corresponding the
calculations with and without \GGp term in the FSI factor
differ substantially. The difference is connected
with allowance for the incoherent rescatterings in the
case of inclusive $(e,e'p)$ reaction. However as we see
from our numerical results the contribution from the
incoherent rescatterings become important only at
sufficiently large missing momenta. It means that
the nuclear transparency measured in a certain restricted  kinematical
region $D$ may considerably differ from the integral nuclear
transparency even if in both cases the inclusive
experimental conditions are imposed. Of course the
fact that the contribution from the incoherent rescatterings
predominantly comes from high- $|\vec{p}_{m}|$ region
does not mean automatically that the nuclear transparency
measured in the kinematical region including sufficiently
small $|\vec{p}_{m}|$ will be close to $T_{A}^{exc}$.
As we have seen even if \GGp term is not taken into account
the missing momentum distribution will considerably differ from the
PWIA case.

The above is
especially important for experimentally disentangling CT effects
in $(e,e'p)$ scattering.
It is clear that definitive conclusions on their role can only
be made if one compares the experimental nuclear transparency
with the one obtained in the Glauber model for the same kinematical
domain $D$. In order to demonstrate the dependence of the
nuclear transparency on the choice of kinematical domain $D$
in the definition (\ref{eq:1.1}) in Fig.~9,~10 we present the results
in the case of the version with \GGp term (solid lines) for
four different windows in the missing momentum.
For the purpose of the comparison we also show in these figures
the inclusive and exclusive integral nuclear transparency computed
using
$\sigma_{in}(pN)$ (long-dashed lines) and $\sigma_{tot}(pN)$
(dash-dotted lines).
 To illustrate the dependence of the Glauber model predictions
on the parameterization of the one-body density matrix, we show
in Fig.~9,~10 the results obtained using the LDA (\ref{eq:2.32}) for
the
one-body density matrix (short-dashed line). The results
for shell model one-body density matrix and its LDA parameterization
differ substantially, and this difference varies with the
missing momentum window. As one can see from Fig.~9,~10
the Glauber theory results
with full shell model density matrix for considered
kinematical windows do considerably
differ from the both integral nuclear transparencies
$T_{A}^{inc}$   and $T_{A}^{exc}$. Important finding is
that despite the increase of $\sigma_{in}(pN)$ for all considered
kinematical domains the nuclear transparency slightly
rises with $Q^{2}$.
We wish specially emphasize sensitivity of $T_{A}(D)$ to the
missing momentum window. One could expect that intranuclear
attenuation can neither be stronger than given by $\sigma_{tot}$
nor weaker than given by $\sigma_{in}$, hence naively
$$T_{A}^{inc}>T_{A}(D)>T_{A}^{tot}$$
The results shown in Fig.~9,~10 clearly demonstrate, that
quantum mechanical distortion effects do not amount to a naive
attenuation. Namely, even at small $\vec{p}_{m}$, where
incoherent elastic rescattering effects are still small
one can easily find $T_{A}(D)>T_{A}^{inc}$.

The kinematical region
$D=(p_{m\perp}<250,~|p_{m,z}|<50$ MeV/c) approximately corresponds
to the kinematical conditions of the recent NE18 experiment
\cite{NE18}. In Fig.~11 we compare the
experimental data \cite{NE18} for $^{12}C$ and $^{56}Fe$ with
the Glauber model predictions.
$^{12}C$ and $^{56}Fe$ are not closed-shell nuclei, $T_{A}(D)$
for these nuclei were calculated assuming that they interpolate
between $T_{A}^{exc}$ and $T_{A}^{inc}$ as for the closed-shell
$^{16}O$ and $^{40}Ca$ nuclei, respectively.
$T_{A}^{inc}$ and $T_{A}^{exc}$
for $^{12}C$ were computed  using the parameterization of the nuclear
density in a form of a sum of Gaussians \cite{Atdata}. In the case of
$^{56}Fe$
the three-parameter Gaussian model \cite{Atdata} was
used. The difference between the charge density distribution
and the proton density distribution was taken into account.
Strong dependence of $T_{A}(D)$ on the missing momentum window
$D$, see Fig.~9,~10, makes the full quantum mechanical treatment of
distortions imperative for a quantitative comparison between the
theory and experiment. This effect is missed in all the previous
calculations of $T_{A}$, which were reviewed to much detail by Makins
and Milner \cite{Milner}. The only exception is paper
\cite{JNE18},
which discussed how $T_{A}(D_{NE18})$ interpolates between
$T_{A}^{exc}$ and $T_{A}^{inc}$, but the analysis \cite{JNE18}
only included the dominant distortion of the transverse momentum
distribution from the incoherent rescatterings. In Fig.~11 we also
show the estimate \cite{onsetCT} for CT effects.

Our values of $T_{A}(D_{NE18})$ are somewhat below the NE18
determinations. To this end, we wish to remind that the NE18
analysis \cite{NE18} uses certain model evaluations of the
denominator in Eq. (\ref{eq:1.1}). One of the key assumptions of
is that, modulo to the overall normalization, the nuclear spectral
function is identical to the PWIA spectral function. Our results
show, that  this can not be correct for the FSI effects, but
the accuracy of the NE18 experiment is not sufficiently high
as to unravel the distortions of the size found in our analysis.
Furthermore, the NE18 analysis introduces renormalization of
$T_{A}$ by the factor $1.11\pm 0.03$ for $^{12}C$ and
$1.22\pm 0.06$ for $^{56}Fe$ nuclei, which renormalization is meant
to account for the missing strength associated with the large-$p_{m}$
component of the spectral function coming from short range
$NN$ correlations. Similar correction for the missing strength
must be included, both in the numerator and denominator
of Eq. (\ref{eq:1.1}), in our analysis too. The recent work
\cite{He4} found strong interference between, and similar
strength of, the FSI and short range correlation effects, which
make the corrections for the $NN$ correlations to the numerator
and denominator different. For this reason the above cited
renormalization effects must be regarded as an indication of the
accuracy of shell model calculations of $T_{A}(D_{NE18})$.
Calculations with $NN$ correlations and correct treatment of FSI are
required for the higher accuracy comparison of the experimental and
theoretical values of $T_{A}$.

The difference between the
theoretical predictions obtained using the full shell model
one-body density matrix and the density matrix of the LDA
(\ref{eq:2.32}) shown in Fig.~9,~10 shows that in the
case
of three-dimensional missing momentum distribution the FSI effects in
quasielastic $(e,e'p)$ scattering is rather sensitive to the
non-diagonal elements of the one-body density matrix. To gain more
insight into the sensitivity to the one-body density matrix, in
Fig.~12 we
compare the results for the $p_{m,z}$ integrated  nuclear transparency
as a function  of $p_{\perp}$ for the full shell model density
matrix and the LDA density matrix (the solid and long-dashed
curves) at $Q^{2}=2$ GeV$^{2}$. The difference of $p_{m,z}$ integrated
transparencies is very large at small $p_{m\perp}$, reaching
$\sim 20$ \% for $^{40}Ca$ nucleus, the full shell model also
predicts much deeper minimum of transparency at $p_{m\perp}\sim 225$
MeV/c, beyond the crossover at $p_{m\perp}\sim 150$ MeV/c.

The difference between the Glauber model predictions for
two versions of the one-body density matrix become more
clear if one compares the results in the case of the non-integrated
nuclear transparency. In Fig.~13,~14 we perform this comparison
for the transverse kinematics. The results for the parallel kinematics
presented in Fig.~15,~16. As one
can see from Fig.~13-16, using the LDA (\ref{eq:2.32}) leads to
an underestimation
of the nuclear transparency at small $|\vec{p}_{m}|$ and
to an overestimation in high-$|\vec{p}_{m}|$  region.
At $\vec{p}_{m}\sim 0$
predictions from full shell model and from LDA differ by $\gsim 20$\%.

One could expect that the LDA (\ref{eq:2.32}) will become more
reliable with increasing of $A$. Our important finding is that the
discrepancy between the Glauber model predictions for two versions
of the one-body density matrix does not reveal any tendency to
disappear with increasing of the nucleus mass number.
This fact once more emphasizes that the quantum interference effects
play important role in the FSI of the struck proton in the
nuclear medium. The failure of the LDA even for sufficiently
heavy nucleus $^{40}Ca$ at the qualitative level can be
explained by the large contribution to the cross section
of $(e,e'p)$ scattering from the events corresponding to
the ejecting of the proton from the nucleus surface.
Evidently there are no any reasons to expect that the
LDA will be reliable in the surface region.
\section{Conclusions}
The purpose of this work has been a study of
the missing momentum distribution in
quasielastic $(e,e'p)$ scattering in the region
of moderate missing momenta , $|\vec{p}_{m}|\lsim 300$ MeV/c,
and high energy, $Q^{2}\sim 2-10$ GeV$^{2}$,
within the Glauber multiple scattering theory.
To perform such an analysis, we generalized the Glauber theory
developed for the hadron-nucleus collisions at high energy,
to the case of $(e,e'p)$ reaction. We presented for the first
time a consistent treatment of the novel effect of interaction
between the two trajectories which enter the calculation of
the FSI-modified one body density matrix and have an origin
in the incoherent elastic rescatterings of the struck proton.

Our numerical results show that the missing momentum distribution
in $(e,e'p)$ scattering are substantially affected by FSI
effects as compared to the PWIA case both for the inclusive
and exclusive conditions. In the studied kinematical region the
distortion effects connected with coherent rescatterings of the
struck proton dominate at $\vec{p}_{m}\lsim 200$ MeV/c. The
contribution from the incoherent rescatterings appears to be
important at $\vec{p}_{m}\gsim 200$ MeV/c.  Our important finding
is that apart from the transverse missing
momentum distribution, incoherent rescattering do
substantially affect also the longitudinal momentum distributions
at high missing momentum. This distortion of longitudinal momentum
distribution is of a purely
quantum-mechanical origin.

Our calculations show that the forward-backward asymmetry
connected with the elastic (diagonal) rescatterings of the
struck proton on the spectator nucleons is larger than expected
from the CT effects by the factor about 2-4 at $Q^{2}\sim 5-10$
 GeV$^{2}$. In the region $Q^{2}\sim 2-3$ GeV$^{2}$ the expected role
of the CT effects is negligible. Thus our results make it clear that
the forward-backward asymmetry practically
can not be considered as a clean signal of the onset CT in the
CEBAF kinematical region.

Using computed three-dimensional missing momentum distribution
we studied the energy dependence of the nuclear transparency for
a few kinematical domains. Our results show that
despite the rise of $\sigma_{in}(pN)$, leading to the
decreasing integral nuclear transparency, the nuclear
transparency for the kinematical domains with $\vec{p}_{m}\lsim 250$
MeV/c even slightly increases with $Q^{2}$. For the first time
we performed the comparison of the Glauber model prediction
with the recent data from NE18 experiment \cite{NE18} accurately
taking into account the kinematical restrictions in the missing
momentum. The energy dependence obtained in the present paper
is close to the one observed in \cite{NE18}. Our detailed
calculations of distortion effects also allowed to highlight
a limited applicability of treatments of FSI effects based on
the conventional DWIA and local-density
approximations.

Our important observation is that in the case of $(e,e'p)$ reaction
the Glauber formalism is incomplete at
sufficiently high longitudinal missing momenta.
We have shown that the standard
Glauber model Ansatz for the attenuation factor leads
to the anomalously slow decrease ( $\propto |p_{m,z}|^{-2}$)
of the missing momentum distribution at high longitudinal
missing momenta. Such a tail is an artifact of neglecting
the finite longitudinal
size of the region where the interaction of the struck
proton with the spectator nucleon takes place.
Taking into account of the finite interaction size
must drastically change the Glauber model predictions
at $|p_{m,z}|\gsim 500$ MeV/c. We checked that
corrections
to the predictions of the standard Glauber approach are still
negligible at $|p_{m,z}|\lsim 300$ MeV/c.
   It is important that the same incompleteness is inherent to, and
persists, also in the color transparency regime at high $Q^{2}$,
where the Glauber theory must be complemented by the off-diagonal
transitions.

The sensitivity of the FSI effects to the finite longitudinal size
of the interaction zone for $pN$ collision has important
implications for interpretation of the experimental
data on $(e,e'p)$ scattering in terms of the short range $NN$
correlations in nuclei. Specifically, it makes it clear that besides
the short range $NN$ correlation the measured missing
momentum distribution at high missing momenta is sensitive
to the nucleon structure as well.
 The results of the detailed analysis
of the influence of the finite nucleon size upon the
missing momentum distribution at high missing momenta
will be presented elsewhere.
\bigskip\\
{\bf Acknowledgements:} Discussions with  S.Fantoni and O.Benhar
are gratefully acknowledged. B.G.Z. thanks S.Fantoni for the
hospitality extended to him
at the Interdisciplinary Laboratory,
SISSA.
This work was partly supported by the
Grant N9S000 from the International Science Foundation and
the INTAS grant 93-239.

\pagebreak

\pagebreak
{\large \bf Figure captions:}
\begin{itemize}

\item[Fig.~1]

{}~- The $Q^{2}$-dependence of nuclear transparency for
$^{16}O(e,e'p)$ and $^{40}Ca(e,e'p)$ scattering.
The long-dashed curve is for the inclusive,
$(p_{m\perp},p_{m,z})$-integrated transparency $T_{A}^{inc}$
as given by Eq.~(\ref{eq:3.8}), the dot-dashed curve is for
the exclusive transprarency as given by Eq.~(\ref{eq:3.9}).

\item[Fig.~2]

{}~- Angular dependence of the inclusive missing momentum distribution
$n_{eff}(p_{m},\theta)$ in $^{16}O(e,e'p)$ scattering, calculated
for FSI without (the left hand boxes) and
including (the right hand boxes) the $\Gamma^{*}\Gamma$ terms,
as compared to the single-particle
momentum distribution $n_{F}(p_{m})$: the short-dashed, solid,
dot-dashed and long-dashed curves are for missing momentum
$p_{m}=150,~200,~250$ and 300~MeV/c, respectively.

\item[Fig.~3]

{}~- The same as Fig.~2, but for the $^{40}Ca(e,e'p)$ scattering.

\item[Fig.~4]

{}~- The missing momentum dependence of nuclear transparency in
parallel kinematics ($\theta=0^{o},~180^{o}$) and transverse
kinematics ($\theta=90^{o}$) calculated (solid curve) for full
FSI, including and (long-dashed curve) in the optical approximation,
not including the $\Gamma^{*}\Gamma$ terms.

\item[Fig.~5]

{}~- Nuclear transparency for $^{40}Ca(e,e'p)$ scattering as a
function of the transverse missing momentum $p_{m\perp}$ at
different fixed values of the longitudinal missing momentum
$p_{m,z}$. The boxes ${\bf (a),(c)}$ are for the full FSI
including the $\Gamma^{*}\Gamma$ terms, the boxes ${\bf (b),(d)}$
are for the optical approximation to FSI, with
the $\Gamma^{*}\Gamma$ terms not included.

\item[Fig.~6]

{}~- Multiple-elastic rescattering decomposition of nuclear
transparency in $^{16}O(e,e'p)$ scattering,
integrated over the longitudinal missing momentum.
The dot-dashed curve is for the exclusive transparency ($\nu=0$),
the solid curve shows the inclusive transparency summed over all
rescatterings (all $\nu$), the dashed curve shows the effect of
including the first elastic rescattering ($\nu=0,~1)$.

\item[Fig.~7]

{}~- The same as Fig.~6, but for $^{40}Ca(e,e'p)$ scattering.

\item[Fig.~8]

{}~- The longitudinal-missing momentum dependence of nuclear
transparency integrated over the transverse missing momentum
$p_{m\perp}$. The solid curves are
for the full FSI
including the $\Gamma^{*}\Gamma$ terms, the long-dashed curves
are for the optical approximation to FSI, with
the $\Gamma^{*}\Gamma$ terms not included.

\item[Fig.~9]

{}~- The $Q^{2}$-dependence of nuclear transparency for
$^{16}O(e,e'p)$ scattering at different windows $D$ in the
transverse $p_{m\perp}$ and longitudinal $p_{m,z}$ missing
momentum in comparison with (the long-dashed curve) the
inclusive transparency $T_{A}^{inc}$ and (the dot-dashed curve)
the exclusive transprarency. The solid curve shows transparency
$T_{A}(D)$, calculated with full treatment of FSI ($\Gamma^{*}
\Gamma$ terms included), for the
$(p_{m\perp},p_{m,z})$-window $D$ as shown in the corresponding
box. The short-dashed curve is the same as the solid curve,
but for the optical approximation description of FSI ($\Gamma^{*}
\Gamma$ terms not included).

\item[Fig.~10]

{}~- The same as Fig.~9, but for the $^{40}Ca(e,e'p)$ scattering.

\item[Fig.~11]

{}~-
Predictions of nuclear transparency for
the missing
momentum window ($p_{m\perp}<250$ Mev/c,$|p_{m,z}|<50$ Mev/c)
in comparison with the NE18 determinations for $^{12}C$
(solid curve) and $^{56}Fe$ (dot-dashed curve) nuclei. For
the $^{12}C$ nucleus we also show the effect of color
transparency (dashed curve) as evaluated in \cite{JNE18}.

\item[Fig.~12]

{}~- The transverse missing-momentum dependence of nuclear
transparency integrated over the longitudinal missing momentum
$p_{m,z}$,
for the full shell-model calculation with the $\Gamma^{*}\Gamma$
terms included (solid curve) and not included (long-dashed curve),
and for the local-density approximation with the $\Gamma^{*}\Gamma$
terms included (short-dashed curve) and not included (dot-dashed
curve). \\
The bottom boxes show the same curves in the blown-up scale.

\item[Fig.~13]

{}~- Nuclear transparency in $^{16}O(e,e'p)$ scattering
in transverse kinematics $p_{m,z}=0$
for the full shell-model calculation with the $\Gamma^{*}\Gamma$
terms included (solid curve) and not included (long-dashed curve),
and for the local-density approximation with the $\Gamma^{*}\Gamma$
terms included (short-dashed curve) and not included (dot-dashed
curve).

\item[Fig.~14]

{}~- The same as Fig.~13, but for $^{40}Ca(e,e'p)$ scattering.

\item[Fig.~15]

{}~- Nuclear transparency in $^{16}O(e,e'p)$ scattering
in parallel kinematics $p_{m\perp}=0$
for the full shell-model calculation with the $\Gamma^{*}\Gamma$
terms included (solid curve) and not included (long-dashed curve),
and for the local-density approximation with the $\Gamma^{*}\Gamma$
terms included (short-dashed curve) and not included (dot-dashed
curve).

\item[Fig.~16]

{}~- The same as Fig.~15, but for $^{40}Ca(e,e'p)$ scattering.
\end{itemize}
\end{document}